\documentclass[aps,prb,amsmath,amssymb,nobibnotes,longbibliography,superscriptaddress,twocolumn]{revtex4-2}

\usepackage{amsmath}
\usepackage{amssymb}
\usepackage{graphicx}
\usepackage{dcolumn}
\usepackage{bm}
\usepackage{bbm}
\usepackage{dsfont}

\usepackage[utf8]{inputenc}
\usepackage[T1]{fontenc}
\usepackage{textgreek}
\usepackage{upgreek}
\usepackage{float}
\usepackage{hyperref}
\usepackage{placeins}
\usepackage{newtxtext}
\usepackage{newtxmath}

\usepackage{comment}

\usepackage[colorinlistoftodos]{todonotes}
\presetkeys{todonotes}{inline}{}

\usepackage{color}
\usepackage[normalem]{ulem}
\usepackage{xcolor}
\definecolor{darkgreen}{RGB}{0 100 0}

\usepackage{natbib}

\newcommand{\ti}[1]{_\text{#1}}

\newcommand{\beginsupplement}{%
        \setcounter{table}{0}
        \renewcommand{\thetable}{S\arabic{table}}%
        \setcounter{figure}{0}
        \renewcommand{\thefigure}{S\arabic{figure}}%
				\renewcommand{\theequation}{S.\arabic{equation}}
     }

\usepackage{soul}

\begin{document}


\title[HgTe CPR]{Giant anomalous Josephson effect as a probe  of spin texture in topological insulators}

\author{Niklas~Hüttner}
\affiliation{Institut f\"ur Experimentelle und Angewandte Physik, University of Regensburg, 93040 Regensburg, Germany}
\author{Andreas Costa}
\affiliation{Institut f\"ur Theoretische Physik, University of Regensburg, 93040 Regensburg, Germany}
\author{Leandro Tosi}
\affiliation{Grupo de Circuitos Cuánticos Bariloche, CAB-CNEA, CONICET and Instituto Balseiro, (8400) San Carlos de Bariloche, Argentina}
\author{Michael Barth}
\affiliation{Institut f\"ur Theoretische Physik, University of Regensburg, 93040 Regensburg, Germany}
\author{Wolfgang Himmler}
\author{Dmitriy A. Kozlov}
\affiliation{Institut f\"ur Experimentelle und Angewandte Physik, University of Regensburg, 93040 Regensburg, Germany}
\author{Leonid Golub}
\affiliation{Institut f\"ur Theoretische Physik, University of Regensburg, 93040 Regensburg, Germany}
\affiliation{Halle-Berlin-Regensburg Cluster of Excellence CCE, University of Regensburg, 93040 Regensburg, Germany}
\author{Nikolay\,N.\,Mikhailov}
\affiliation{Rzhanov Institute of Semiconductor Physics, 630090 Novosibirsk, Russian Federation}
\author{Klaus Richter}
\affiliation{Institut f\"ur Theoretische Physik, University of Regensburg, 93040 Regensburg, Germany}
\author{Dieter Weiss}

\author{Christoph Strunk}
\affiliation{Institut f\"ur Experimentelle und Angewandte Physik, University of Regensburg, 93040 Regensburg, Germany}
\affiliation{Halle-Berlin-Regensburg Cluster of Excellence CCE, University of Regensburg, 93040 Regensburg, Germany}
\author{Nicola Paradiso}
\affiliation{Institut f\"ur Experimentelle und Angewandte Physik, University of Regensburg, 93040 Regensburg, Germany}

\date{\today}

\begin{abstract}	
Surface states of topological insulators feature chiral spin-momentum locking. When such states are used as weak link between two superconductors, their spin texture gives rise to the anomalous Josephson effect, i.e., to a $\varphi_0$ shift in the current phase relation. In this work, we explore the anomalous Josephson effect in junctions where the weak link is a HgTe nanowire. We observe a giant anomalous $\varphi_0$-shift of the current-phase relation, which we attribute to the fact that HgTe surface states feature a \textit{single} Fermi contour. Moreover, by varying the orientation of the in-plane magnetic field, we obtain information about the spin texture in momentum space. In particular, we found that the spin is not exactly perpendicular to the momentum, but shows a significant deviation of 19 degrees. Our results establish the anomalous Josephson effect as a sensitive tool to probe the spin texture of chiral 2D systems. 
\end{abstract}

\maketitle

Josephson junctions are formed by a weak coupling of two superconducting electrodes and characterized by the flow of a dissipationless current of Cooper pairs~\cite{Josephson1962,Josephson1964}. The supercurrent is determined by the phase difference between the superconductors and by the properties of the weak link~\cite{Likharev1979}. For mesoscopic weak links, the dispersion relation close to the Fermi level determines the velocity of the electrons and holes contributing to the formation of Andreev bound states (ABS)~\cite{Kulik1969a}. Since these states carry the supercurrent, the Josephson effect has proven to be a powerful tool to characterize a broad variety of weak links, such as atomic-size metallic contacts~\cite{DellaRocca2007}, carbon nanotubes~\cite{Pillet2010}, graphene~\cite{Bretheau2017}, quasi-one-dimensional semiconducting wires~\cite{Prada2020}, and two-dimensional electron gases~\cite{Irie2016}. 

When the dispersion relation of the weak link has a spin texture, this is inherited by the ABS and can lead to interesting spin-dependent properties~\cite{Fu2009,Beenakker2013,Potter2013}, particularly in combination with a Zeeman energy. Depending on the spin orientation with respect to the momentum, an applied magnetic field may lead to a phase offset $\varphi_0$ in the current-phase relation (CPR), giving rise to the so-called anomalous Josephson effect~\cite{Buzdin2008,Reynoso2008,Yokoyama2014,Dolcini2015}. Such $\varphi_0$-junctions have been reported in both 1D~\cite{Szombati2016,Strambini2020,FrolovQW2022} and 2D~\cite{Mayer2020b,Dartiailh2021,Haxell2023,Reinhardt2024} systems with strong Rashba spin–orbit coupling (SOC). In 2D systems, the ratio between $\varphi_0$ and the applied magnetic field $B$ is at most on the order of $\partial \varphi_0/\partial B \approx 2\pi$~rad/T. Anomalous phase shifts have also been reported for weak links based on the topological insulator~(TI) Bi$_2$Se$_3$, with responses of the order of $\partial \varphi_0/\partial B \approx 10\pi$~rad/T \cite{Assouline2019}.

Because of their ballistic character, the surface states of the strained 3D TI HgTe offer an excellent playground to investigate the impact of spin textures on the Josephson supercurrent. Thanks to the strong spin-orbit interaction and the band inversion in the bulk, devices based on this material have served as a unique platform to explore novel physical systems for many years: the 2D TI~\cite{Konig2007} and its more exotic counterpart, the topological Anderson insulator~\cite{Khudaiberdiev2025}; topological superconductivity~\cite{Hart2014,Fischer2022}, and Majorana physics \cite{Fu2008,Bocquillon2017}. 

In HgTe-nanowire-based Josephson junctions, the supercurrent was shown to be carried by surface states that form ABS~\cite{Fischer2022}. 
Many properties of these surface states are fairly well understood. In particular, the spatial separation of the wave functions on opposite surfaces was first demonstrated in HgTe thin films~\cite{Kozlov2016} and later in HgTe nanowires~\cite{Ziegler2018}, where the closed-loop geometry wrapping around the nanowire enables the observation of Aharonov–Bohm type oscillations in a parallel magnetic field. However, relatively little is known about the spin texture of the surface states. The absence of spin degeneracy has been experimentally confirmed~\cite{Maier2017,Koop2024}, and the observation of weak anti-localization supports the theoretically predicted strong coupling between the spin orientation and momentum in these states~\cite{Savchenko2016,Savchenko2021}. Spin-resolved ARPES is unlikely to shed much additional light on this question. The main challenge lies in the small bulk energy gap of strained HgTe ($\approx$15~meV), which makes the accessible energy range of surface states in experiments much smaller than the typical ARPES detection window (on the order of 1~eV)~\cite{Bruene2011, crauste2013topologicalsurfacestatesstrained}.
Therefore, detailed experimental evidence for the spin texture of the topological surface states in HgTe films is still lacking. 

Here, we present measurements of the current-phase relation of Josephson junctions with HgTe as weak link. 
We use the strained 88~nm thick HgTe film, which is known as a strong 3D TI~\cite{Fu2007}.
When an in-plane magnetic field perpendicular to the wire axis is applied, a giant anomalous phase 
is observed. This response is much larger than the shift reported in Rashba weak links. We attribute this to both the larger $g$-factor and to the 
topological nature of electrons in HgTe with Dirac-like dispersion. 
Moreover, we make use of the angle dependence of the $\varphi_0$-shift to probe the surface-state spin texture, namely, the relative orientation of the spin with respect to the momentum.

\begin{figure*}[t!]
\begin{center}
\includegraphics[width=\textwidth]{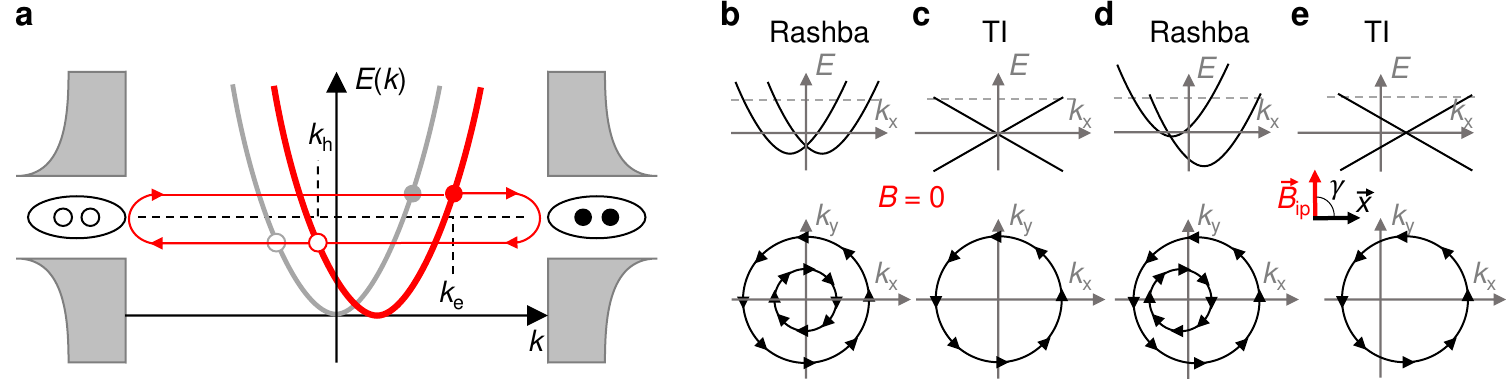}
\end{center}
\caption{\textbf{a}, Scheme of supercurrent transport mediated by Andreev bound states. The graph in gray (red) shows the situation where counterpropagating quasiparticles have the same (different) momentum, namely, $k_e = k_h$ ($k_e \neq k_h$). For finite accumulated phase $(k_e-k_h)L$ over the junction length $L$, the current-phase relation is shifted by an anomalous phase $\varphi_0$ and it is no longer antisymmetric [$I(\varphi) \neq -I(-\varphi)$].
\textbf{b-e}, 
Dispersion relation (top) and Fermi surfaces (bottom) for a parabolic (\textbf{b,d}) 2D electron system and a Dirac cone of a single surface of a 3D TI (\textbf{c,e}).
The arrows in the Fermi surfaces indicate the spin direction with respect to the momentum, which in the absence of bulk-inversion asymmetry (as for the sketched cases) is  $\gamma=90^{\circ}$. If the spin is locked perpendicular to the momentum, then an in-plane field $\vec{B}$ directed along $y$ (\textbf{d,e}) shifts the Fermi contours along the current direction $\hat{x}$, as in panel \textbf{a}, inducing a $\varphi_0$-shift. 
}
\label{fig:idea}
\end{figure*}

As illustrated in Fig.~\ref{fig:idea}\textbf{a}, the Cooper pairs transferred in the Josephson effect can be described in terms of ABS. These are electron-hole excitations confined in the weak link: for every round trip within the weak link, the quasiparticles must acquire a phase of 2$\pi$. This phase is given by: (i) the phase acquired during propagation $ (k_e - k_h) L$, where $k_e$ ($k_h$) is the $k$-vector of the electron (hole) and $L$ the length of the weak link; (ii) the phase due to Andreev reflection, which depends on the energy $E$ and the phase difference $\varphi$ between the superconducting electrodes. 

In addition, normal scattering in the weak link and at the interfaces with the superconductors also has to be taken into account. For a single channel in the short-junction limit $L\ll \xi_0$, with $\xi_0=\hbar v_F/\pi \Delta$ being the coherence length (for Fermi velocity $v_F$ and superconducting gap $\Delta$), a pair of ABS is obtained with energies $E(\varphi)=\pm\Delta\sqrt{1-\tau \sin^2\left(\varphi/2\right)}$, where $\tau$ is the transmission probability~\cite{Haberkorn1978,Furusaki_1991,BeenakkerPRL91,RMPGolubov}. The supercurrent can be obtained as $I(\varphi)=\frac{1}{\phi_0}\partial E/\partial \varphi$, where $\phi_0=\frac{\Phi_0}{2\pi}$ is the reduced flux quantum, yielding 
\begin{equation}
I(\varphi,T)=\frac{\Delta \tau \sin\varphi\tanh\left(\frac{E(\varphi)}{2k_B T} \right)}{2 \phi_0 \sqrt{1-\tau \sin^2\left(\varphi/2\right)}} ,
    \label{eq:furusaki}
\end{equation}
at finite temperature $T$ \cite{Furusaki_1991, BeenakkerPRL91}. 

ABS provide a simple picture for the emergence of the anomalous Josephson effect. In a nutshell, the $\varphi_0$-shift is obtained when right-moving and left-moving quasiparticles in the weak link propagate with different $k$-vectors~\cite{Yokoyama2014}. In weak links with strong Rashba SOC, this is obtained by applying an in-plane field perpendicular to the current direction (see Figs. \ref{fig:idea}\textbf{b,d}), which shifts the two circular Fermi contours in opposite directions. If we consider only $k$-vectors along the $x$-direction (1D limit), the $\varphi_0$-shift is expected to be~\cite{Yokoyama2014}
\begin{equation}
    \varphi_0=-L(k_{F,+}^>+k_{F,+}^<+k_{F,-}^>+k_{F,-}^<)/2,
    \label{eq:phi0main}
\end{equation}
where $>$ ($<$) indicates positive (negative) $k_x$ and $+$ ($-$) indicates $k_F$ belonging to the parabolic dispersion shifted towards the right (left) direction \cite{Yokoyama2014} (see Fig. \ref{fig:idea}\textbf{d}). In strictly 1D systems with ideal parabolic dispersion, the shift of the outer terms ($k_{F,+}^>,k_{F,-}^<$) is equal and opposite in sign to that of the inner terms ($k_{F,+}^<,k_{F,-}^>$). In quasi-1D, a finite $\varphi_0$ is nevertheless expected owing to the mixing between the subbands~\cite{Yokoyama2014}. In 2D systems, the $\varphi_0$ is also deduced from the different density of states between the outer and inner Fermi contours. 
The key idea at the basis of the present work is that the opposite shift of the inner Fermi contour mostly cancels the effect due to the shift of the outer one. Instead, without the inner Fermi contour, one would expect a much larger anomalous shift in terms of $\Delta \varphi_0/\Delta B_y$. Our goal is to show such a dramatic enhancement of $|\varphi_0|$ for the surface states of a 3D TI with a single circular Fermi contour featuring a Dirac-like dispersion and well-defined spin-momentum locking (see Figs.~\ref{fig:idea}\textbf{c,e}). 
The angle $\theta$ between momentum vector and spin determines the Zeeman-field direction $\gamma$ that maximizes the difference between $k_{F}^>$ and $k_{F}^<$ and thus the $\varphi_0$-shift. For example, 2D electron gases with parabolic dispersion and Rashba SOC have $\theta=90^{\circ}$, so that the largest magnetochiral effects are obtained by applying an in-plane field perpendicular to the current~\cite{Reinhardt2024} (see Figs.~\ref{fig:idea} \textbf{b,d}). One can show~\cite{Kang2024} that for any other $\theta$, the same phenomenology is obtained by accordingly rotating the field in such a way to be parallel to the spin corresponding to the momentum along the current.

\begin{figure*}[t!]
\begin{center}
\includegraphics[width=\textwidth]{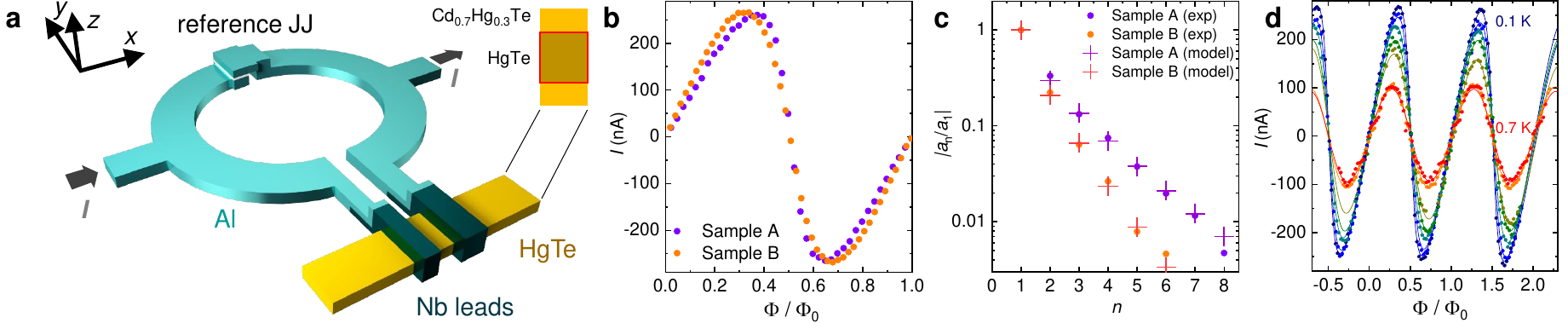}
\end{center}
\caption{\textbf{a}, Scheme of our asymmetric SQUID device consisting of a reference Al/AlOx/Al tunnel junction (top) and a Nb/HgTe/Nb SNS junction (bottom). A magnetic field applied along the $z$-direction, $B_z$, produces the flux $\Phi$, which modulates the critical current of the SQUID. The red rectangle indicates the surface states wrapping around the HgTe wire. 
\textbf{b},~Measured current-phase relation (CPR) of the HgTe junction for two distinct samples (see text). 
\textbf{c}, Modulus of the $n$-th Fourier components (dots) for the two CPRs in \textbf{b}. The exponential decay agrees with the Furusaki-Beenakker predictions (Eq.~\ref{eq:furusaki}, crosses) in the 1D limit and makes it possible to deduce the average transmission $\tau$.
\textbf{d}, CPRs of sample A measured at different temperatures $T$. When increasing the temperature from 100\,mK to 700\,mK, the relatively stronger suppression of the higher harmonics renders the CPR more sinusoidal. 
}
\label{fig:chara}
\end{figure*}

Our devices are fabricated starting from a stack consisting of a $40\,$nm CdTe capping layer, $30\,$nm Cd$_{0.35}$Hg$_{0.65}$Te, 88\,nm strained HgTe grown along the [013] direction, and again $30\,$nm Cd$_{0.35}$Hg$_{0.65}$Te forming a quantum well on top of a $4\,$µm buffer layer on a GaAs substrate.
HgTe nanowires, with a width on the order of 1~$\mu$m, are defined by electron-beam lithography followed by wet etching. The surface states of the nanowires exhibit high charge-carrier mobilities, typically of several $10^5$\,cm$^2/$V$\cdot$s~\cite{Kozlov2014}.
The weak link is also defined by electron-beam lithography of Nb contacts, deposited after an \textit{in-situ} Ar-ion milling and capped by a Pt layer to prevent oxidation. An asymmetric SQUID is formed by inserting the weak link in a superconducting Al loop with a reference junction, as sketched in Fig.~\ref{fig:chara}\textbf{a}. It consists of an Al-AlOx-Al SIS junction fabricated using the Manhattan technique~\cite{osman2021,chang2025}. We present the results from two samples, labeled A and B, with a slightly different geometry~\cite{supp}.

For our strongly asymmetric SQUID (the critical current of the reference SIS junction is typically 30-100 
times larger than that of the HgTe junction) with negligible loop inductance, the CPR of the small junction can be measured as a modulation of the reference junction's critical current controlled by the flux $\Phi$ enclosed in the loop, provided that the phase $\varphi$ can be associated to $\Phi/\phi_0$. Critical-current measurements are performed in a dilution refrigerator at a base temperature of $\sim40\,$mK unless otherwise specified. 

Figure~\ref{fig:chara}\textbf{b} shows the CPR measured for samples A and B in the absence of an in-plane field. These curves are obtained by measuring a series of 
current-voltage characteristics, typically several hundreds for each $B_z$ value. The average switching current is then recorded as a function of $B_z$. To obtain the graph in Fig.~\ref{fig:chara}\textbf{b}, we subtract the reference junction's critical-current background. 
We notice that the CPR curves are strongly skewed, indicating a high transmission of the ABS~\cite{Sochnikov2015}. After correcting for the small self-inductance (see Supplementary Information~\cite{supp}), the data can be fitted using the Furusaki-Beenakker formula in Eq.~\ref{eq:furusaki}, giving the effective transmissions $\tau_A=0.94\pm 0.01$ and $\tau_B=0.82\pm0.01$. Figure~\ref{fig:chara}\textbf{c} shows the amplitude of the $n$th Fourier component of both CPRs, plotted in a semi-log scale (dots). The exponential decay is in very good agreement with the theory (crosses). According to the Furusaki-Beenakker model, the higher harmonics of the CPR are rapidly suppressed as soon as $T$ becomes a significant fraction of the gap induced in the surface states underneath Nb, $\Delta^{\ast}_\text{A (B)}=150$~\textmu eV (130~\textmu eV). This is observed in Figure~\ref{fig:chara}\textbf{d}, which shows the temperature dependence of the CPR in the range from 100-700~mK, where the CPR evolves from highly skewed to perfectly sinusoidal. Moreover, such a strong temperature dependence further confirms that the skewness does not originate from the SQUID loop inductance $L_l$. Indeed, the $\beta$-factors for the SQUIDs are $\beta_A=2I_0L_{l}/\Phi_0=1.1 \ 10^{-2}$ and $\beta_B=8\ 10^{-3}$, both $\ll 1$ (for further details see Supplemental Material). 

\begin{figure*}[tb]
\begin{center}
\includegraphics[width=\textwidth]{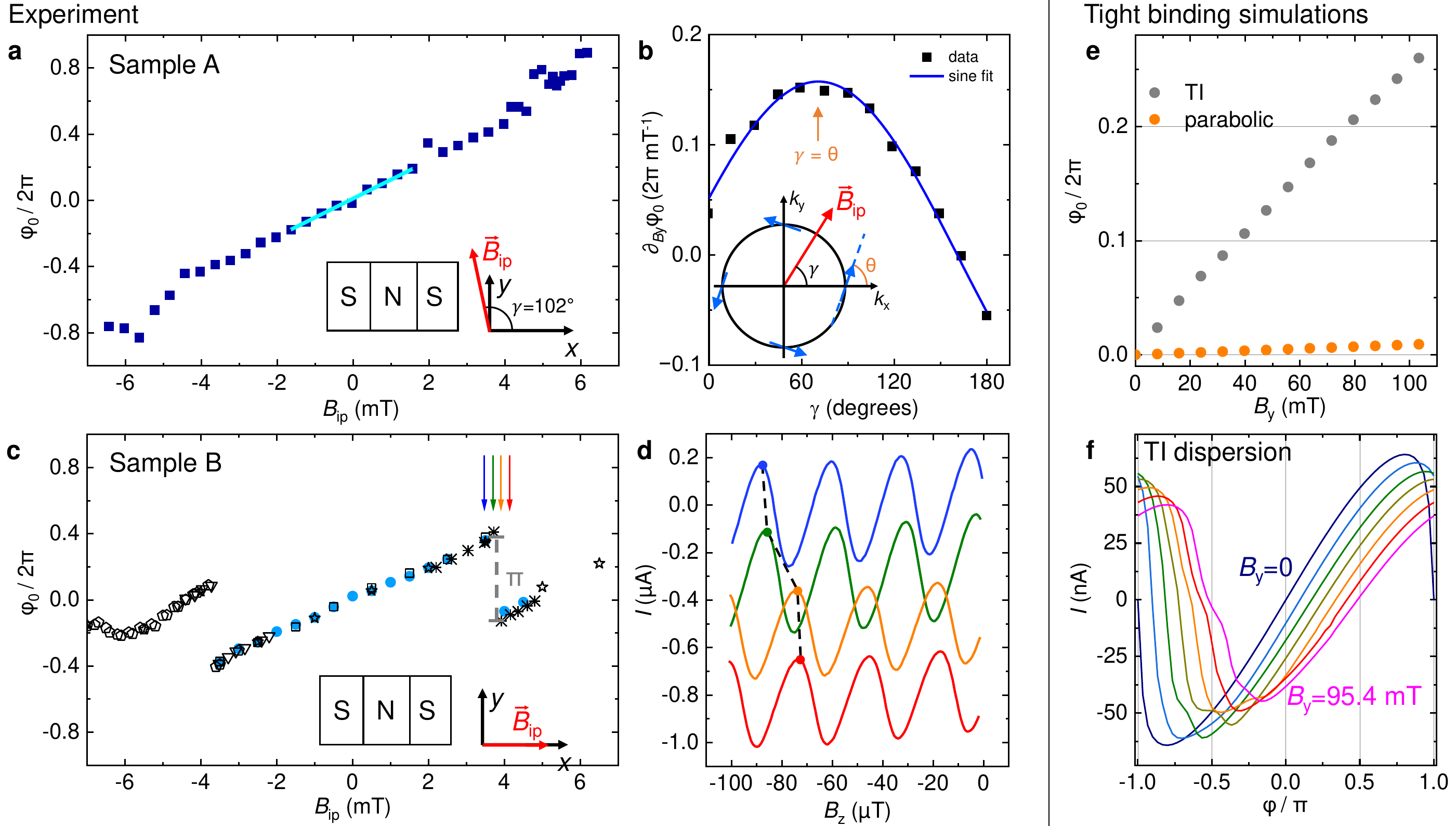}
\end{center}
\caption{\textbf{Anomalous $\varphi_0$-shift as a function of the in-plane field:} \textbf{a}, Anomalous phase shift $\varphi_0$ measured on sample A and normalized to $2\pi$ for the indicated orientation (angle $\gamma =102^{\circ}$) of the in-plane field $\vec{B}_{\text{ip}}$ with respect to the current, namely, the $x$-axis (see inset). \textbf{b}, Slope $\partial \varphi_0/\partial B_y$ measured near $B_y=0$ as a function of the angle $\gamma$, together with a sine fit (fit parameters: horizontal offset $ -19^{\circ}$ and amplitude $0.16\cdot 2\pi$/mT). Data in \textbf{a} and \textbf{b} are obtained from different cool-downs.
\textbf{c}, Anomalous $\varphi_0$-shift measured on sample B for $B_\text{ip}$ parallel to the $y$-axis. The different symbols refer to eight different measurement runs (see text).
\textbf{d}, Exemplary CPRs measured on sample B for four values of $B_\text{ip}$ indicated with arrows in \textbf{c}: $3.48$~mT (blue), $3.72$~mT (green), $3.92$~mT (orange), and $4.16$~mT (red).
\textbf{e}, Tight-binding simulations of the $\varphi_0$-shift in an effectively 2D SNS nanowire junction using a linear TI (gray) or a parabolic (orange) single-electron dispersion and applying the in-plane magnetic field $ B_y $ perpendicular to the current direction; see text for details. 
\textbf{f}, Computed CPRs for the linear TI electronic dispersion and magnetic field $B_y$ ranging from $ 0 $ to $ 95.4 \, \mathrm{mT} $ in steps of $ 15.9 \, \mathrm{mT} $. 
}
\label{fig:phi0}
\end{figure*}

We now focus on the anomalous Josephson effect. To measure $\varphi_0$, it is absolutely crucial to be able to determine the effective zero of the out-of-plane magnetic field with an accuracy much better than the field corresponding to a flux quantum in the SQUID loop (30~$\mu$T in our case). We have thus developed a rigorous out-of-plane field compensation protocol, which we validated by comparison of three independent methods. The routine is repeated every time a new value of the in-plane field is set in order to null out-of-plane field components due to imperfect alignment. An upper bound for the error when nulling the out-of-plane field is 1~µT for the maximum in-plane field applied (6~mT), which corresponds to 3\% of a flux quantum.
The out-of-plane field compensation is the most important and delicate experimental aspect of our measurements. Details can be found in the Supplement~\cite{supp}.

Figure~\ref{fig:phi0} illustrates the two main results of our work, namely, the observation of a giant anomalous shift and the determination of the spin texture orientation for the surface states. 
In Fig.~\ref{fig:phi0}\textbf{a} we show the $\varphi_0$-shift in sample A as a function of in-plane field $\vec{B}_{\text{ip}}$ applied along a direction nearly perpendicular (102$^{\circ}$) to the current (see inset).
Each data point is obtained by measuring the phase shift $\varphi_0$ of the CPR such that $I(\varphi_0)=0$. The CPR is measured over several oscillations. 
The graph shows a $\varphi_0$ shift essentially linear in $B_\text{ip}$. The slope $\partial \varphi_0/\partial B_\text{ip}$ is very large, of the order of $ 110 \cdot 2\pi$~rad/T, which is about 2 orders of magnitude larger than what is typically reported in 2DEGs with parabolic dispersion and large Rashba SOC, as e.g.,~Al/InAs~\cite{Mayer2020b,Reinhardt2024}. 
Our second key result is the dependence of the slope $\partial \varphi_0/\partial B_\text{ip}$  on the in-plane field orientation, which reflects the spin texture at the Fermi contour, as discussed above. To determine the spin-to-momentum angle $\theta$, we repeated the measurement of Fig.~\ref{fig:phi0}\textbf{a} for different orientations of the field, represented by the angle $\gamma$ between $\vec{B}_{\text{ip}}$ and $\hat{x}$ (current axis). For each $\gamma$ value, we obtain $\partial \varphi_0/\partial B_\text{ip}$ by a linear fit near $B_{\text{ip}}=0$ (see cyan line in Fig.~\ref{fig:phi0}\textbf{a}). In the textbook case of parabolic dispersion with Rashba SOC, the angle between spin and momentum is $\theta= 90^{\circ}$, thus the maximum in the magnetochiral response is expected for $\gamma=90^{\circ}$. Data in Fig.~\ref{fig:phi0}\textbf{b} can be fitted with a sine dependence with no vertical offset and with a horizontal shift of $-19^{\circ}\pm 3^{\circ}$, implying $\theta=71^{\circ}\pm 3^{\circ}$ (see Supplemental Material~\cite{supp} for the error determination).
In other words, to obtain the maximum Zeeman shift for electron states with momentum along the current direction ($\hat{x}$), 
a magnetic field must be applied at $-19^\circ$ with respect to the $\hat{y}$-axis, namely, along the direction of electron spin associated with these states.

Figure~\ref{fig:phi0}\textbf{c} shows $\varphi_0(B_\text{ip})$ measured for sample B with $\vec{B}_\text{ip}$ approximately parallel to the current.
The different symbols refer to eight different  sweeps made with different sweep parameters. The results are extremely reproducible within the same cool-down, even for moderate change of temperature; see measurements at $T=40$~mK ($T=200$~mK) indicated with blue (black) symbols. A peculiar feature observed in the measured $\varphi_0(B_\text{ip})$ curves is the presence of sudden jumps in $\varphi_0$. Figure~\ref{fig:phi0}\textbf{d} shows the CPRs measured near one of these jumps (for $B_\text{ip}$ values indicated with arrows in panel \textbf{c}). While $\varphi_0$ varies continuously below and above $B_\text{ip}=3.82$~mT, at that value it jumps almost exactly by $\pi$. It would be tempting to attribute this jump to a Zeeman-driven 0--$\pi$ transition~\cite{Yokoyama2014, Costa2023}. However, all theoretical models predicting such transitions also predict a drastic change in the CPR shape close to the transition point. Instead, in the field regime studied here, the CPR does not change shape (see Fig.~\ref{fig:phi0}\textbf{d}): it just shifts rigidly, with a sudden glitch in correspondence to the phase jump. We considered the possibility of an extrinsic origin of such jumps, as, e.g., the sudden nucleation of vortices. These, however, seem to be incompatible with the accuracy of our compensation routine. A thorough discussion about these checks can be found in the Supplement~\cite{supp}. Moreover, even if Abrikosov vortices nucleate somewhere in the SQUID area, it is not clear how this could produce a jump of the order of $\pi$. In our view, the origin of these jumps is still an open question.

As discussed above, our intuitive explanation for the giant $\varphi_0$-shifts in HgTe weak links is based on the fact that the TI surface states that wrap around the HgTe wire have a single Dirac cone and, consequently, only \textit{one} Fermi surface. This contrasts with the two parabolas and counteracting Fermi surfaces with opposite spin textures in the InAs 2DEGs.
To corroborate this picture, we have performed tight-binding calculations similar to those presented in Ref.~\cite{HimmlerPRR2023}, which successfully described the orbital effects observed for large values of $B_\text{ip}$. In our case, we set the flux over the wire cross-section to zero and consider the field applied exclusively along $\hat{y}$. In this way, we exclude orbital effects e.g., as those discussed in Ref.~\cite{HimmlerPRR2023}, and isolate exclusively the Zeeman contribution. Except for this important difference, our numerical calculations are implemented in a similar way as  those reported in Ref.~\cite{HimmlerPRR2023}:  we consider a minimal model of an effectively 2D-reduced nanowire Josephson junction using the open-source tight-binding package \textsc{Kwant}~\cite{Groth2014}. The details of the model can be found in the Supplement~\cite{supp}. The key result to retain is that the slope, $\partial \varphi_0/\partial B_y$, strongly depends on the type of electron dispersion relation. In the case of TI surface states, the typical electron (hole) Hamiltonian contains the following diagonal terms in the Bogoliubov-de Gennes equation 
\begin{equation}
\label{eq:Hamiltonian}
    \hat{h}_\mathrm{e / h}^\mathrm{TI} = \pm \hbar v_\mathrm{F} \big( - k_x \hat{\sigma}_y + k_y \hat{\sigma}_x \big) \mp \mu \hat{\sigma}_0 \pm U(x) - g \mu_\mathrm{B} B_y \hat{\sigma}_y / 2 , 
\end{equation}
where the $ +(-) $ sign stands for electrons (holes). Here, $ \mu $ is the chemical potential, $ \hat{\sigma}_i$ corresponds to the $ i $th Pauli matrix and $\hat{\sigma}_0 $ is the $ 2 \times 2$ identity matrix. Furthermore, $ U(x) = U_0 \hat{\sigma}_0 [\delta(x) + \delta(x - l_\mathrm{w})] $ describes delta-like barriers at the two interfaces between the weak link of length $l_\mathrm{w}$ and the superconducting leads to model the slightly reduced transparency of the weak link suggested by the measurements.

The calculation of the anomalous shift $\varphi_0$ as a function of $B_y$ is repeated for the single-particle Hamiltonian component typical for 2DEGs with parabolic dispersion and large Rashba SOC $ \alpha $, as e.g., for Al-proximitized InAs quantum wells,   
\begin{multline}
    \hat{h}_\mathrm{e / h}^\mathrm{parabolic} = \pm \hbar v_\mathrm{F} / k_\mathrm{F} (k_x^2 + k_y^2) \hat{\sigma}_0 \mp \mu \hat{\sigma}_0 \pm U(x) - g \mu_\mathrm{B} B_y \hat{\sigma}_y / 2 \\ + \alpha (- k_x \hat{\sigma}_y + k_y \hat{\sigma}_x ).
\end{multline}
The comparison of the resulting $ \varphi_0 (B_y) $ with the TI case (assuming Rashba SOC $ \alpha = 10 \, \mathrm{meV} \, \mathrm{nm} $) is shown in~Fig.~\ref{fig:phi0}\textbf{e}, while in Fig.~\ref{fig:phi0}\textbf{f} we show the corresponding CPRs. 
We notice that the slope $\partial \varphi_0/\partial B_y$ in the case of the TI dispersion is significantly larger than that for parabolic dispersion, which is barely discernible on the same scale. An additional calculation adding an \textit{explicit} Rashba term to the TI Hamiltonian showed a negligible difference, whereas such a term is crucial for observing a $\varphi_0$-shift in parabolic dispersion 2DEGs. This corroborates our interpretation: in the TI case it is the single uncompensated Fermi contour that ultimately produces the large $\varphi_0$ and not the explicit Rashba SOC. However, the computed slope $\partial \varphi_0 /\partial B_y$ is still 45 times smaller than the measured one (notice the different abscissa scale in Figs.~\ref{fig:phi0}\textbf{a} and \textbf{e}). 
This underestimate could be due to the fact that  we used $g=50$, while  the $ g $-factor of HgTe nanowires has been reported to vary significantly under modest gating, e.g., from zero to values above 600~\cite{Reuther2013}.

For simplicity, we have assumed in our calculation that the spin texture is perpendicular to $\vec{k}$ by taking cross-product-like terms $\vec{k}\times \hat{\vec{\sigma}}$ in the Hamiltonian, which implies a spin-to-momentum angle $\theta=90^\circ$. In this case, magnetochiral effects are maximized for magnetic fields oriented perpendicular to the current.  As shown in Ref.~\cite{Kang2024}, in the case of an arbitrary $\theta$-angle, the in-plane field must be rotated accordingly in such a way to be parallel to the spin corresponding to the momentum direction along the current. Except for that, magnetochiral effects will look the same~\cite{Kang2024}.

The computed deviation of the angle $\theta$  from the Rashba value $\theta=90^\circ$ is caused by the low symmetry of the (013) HgTe surface, where no spatial symmetry operations are left except for the identity (point group $C_1$).
Hence, the $ \vec{k} $-linear Hamiltonian has the most generic form~\cite{Ganichev2014}
\begin{equation}
\label{eq:h_generic}
\hat{h}_\mathrm{e / h}^\mathrm{TI, lin} = \pm \hbar \sum_{\nu=x,y,z}\sum_{i=x,y} v_{\nu i} \hat\sigma_\nu {k}_i
\end{equation}
with six linearly-independent coefficients $v_{\nu i}$.
The situation with (013) HgTe quantum wells differs strongly from 2D states on the (001) surface of HgTe, where the SOC arises due to strain~\cite{Kirtschig2016}, and from Bi$_2$Se$_3$ (111) surfaces~\cite{Assouline2019}, where the $ \vec{k} $-linear coupling is the Rashba type only. 

The angle $\theta$ is determined by ratios of the quantities $v_{\nu i}$ in the Hamiltonian~\eqref{eq:h_generic}.
If, in addition to the Rashba term, we take into account the 2D Weyl (or chiral) term $\Delta \hat{h}_\mathrm{e / h}^\mathrm{TI}=\pm \hbar \tilde{v} (\hat\sigma_x k_x + \hat\sigma_y k_y)$ in the Hamiltonian~\eqref{eq:Hamiltonian}, then the dispersion remains linear and isotropic,
but
the spin texture of the eigenstates lies in the surface plane at an angle $\theta=90^\circ - \arctan(\tilde{v}/v_{\rm F})$ to the wavevector $ \vec{k} $. 

In conclusion, we have presented measurements of the anomalous Josephson effect in HgTe-based weak links. We observed a giant Zeeman-field response of the anomalous phase up to $ 150 \cdot 2\pi$~T$^{-1}$, which we attribute to the single spin-textured Fermi surface in HgTe. The anomalous shift thus probes the spin-to-momentum angle $\theta$. Our theoretical analysis reproduces the large $\varphi_0$-shifts of systems with topological Dirac-like dispersion compared to parabolic Rashba dispersion. Beyond their immediate significance for HgTe, our findings serve as a reference point for identifying materials with optimal magnetochiral responses. Furthermore, our method to extract the spin-to-momentum orientation by pure transport measurements offers a versatile tool applicable to a broad class of systems with spin-momentum locking.

\begin{acknowledgments}
We thank Marco Aprili for fruitful discussion.
The work in Regensburg was funded by the Deutsche Forschungsgemeinschaft (DFG, German Research Foundation) within
Project-ID 314695032 – SFB 1277.
N.P. and C.S. acknowledge funding by EU’s HORIZON-RIA Programme under Grant No. 101135240 (JOGATE).
L.T. acknowledges the Georg Forster Fellowship from the Humboldt Foundation, N.H., D.A.K. and D.W. support 
by the European Research
Council (ERC) under the European Union’s Horizon 2020
research and innovation program (Grant Agreement No.
787515, “ProMotion”). 
D.A.K. and L.G. acknowledge funding by the DFG 
via Project-ID 521083032 (Ga501/19).
L.G. was funded by the German Research Foundation (DFG) as part of the German Excellence Strategy – EXC3112/1 – 533767171 (Center for Chiral Electronics).
A.C. acknowledges funding by the DFG 
within Project-ID 454646522. 
\end{acknowledgments}
\vspace{2mm}


\bibliography{biblio}

\clearpage
\newpage
\beginsupplement

\onecolumngrid
\begin{center}
\textbf{\Large Supplementary Information} 

\vspace{0.5cm}

\end{center}
\FloatBarrier

\section{Theoretical model}

 We follow the approach described in Ref.~\cite{HimmlerPRR2023} to implement a minimal toy model of the nanowire Josephson junction in the tight-binding package \textsc{Kwant}~\cite{Groth2014}. The nanowire is unfolded into the two-dimensional $ \hat{x} $--$ \hat{y} $-plane such that the normal link covers the coordinate regions $ 0 \leq x \leq l_\mathrm{w} $ and $ 0 \leq y \leq 2 (w_\mathrm{w} + h) $, where $ 2 (w_\mathrm{w} + h) $ is the circumference of the wire ($ h = 80 \, \mathrm{nm} $ is the height of the HgTe film). 
 
 To compute the CPRs of the nanowire Josephson junction, we attach translationally invariant, semi-infinite superconducting leads with tunable phase difference $ \varphi $ to the weak link and evaluate~\cite{Furusaki1994,Zuo2017}  
\begin{equation}
    I(\varphi) = 2 \frac{e k_\mathrm{B} T}{\hbar} \sum_{n=0}^\infty \sum_{\substack{i \in R \\ j \in L}} \mathrm{Im} \big( H_{ji} G_{ij}^r (\mathrm{i} \omega_n) - H_{ij} G_{ji}^r (\mathrm{i} \omega_n) \big) ;
\end{equation}
$ \omega_n = (2n+1) \pi k_\mathrm{B} T / \hbar $ are the fermionic Matsubara frequencies (we set $ k_\mathrm{B} T \approx 2.6 \, \upmu \mathrm{eV} $ to model ultra-low temperature $ T = 30 \, \mathrm{mK} $), while the labels $ i $ and $ j $ address the lattice sites in two adjacent transverse lattice rows $ R $ and $ L $ (technically, we calculate the current flowing in the junction in the region between these two transverse cuts; as the current is conserved, this is equivalent to the Josephson current). 

The hopping matrix elements from site $ i $ to $ j $, $ H_{ij} $, and off-diagonal elements of the Green's function, $ G_{ij} $, are numerically extracted from the Bogoliubov--de Gennes Hamiltonian of the junction, 
\begin{equation}
    \hat{\mathcal{H}} = \left[ \begin{matrix} \hat{h}_\mathrm{e} & \Delta(x) \\ \Delta^\dagger(x) & \hat{h}_\mathrm{h} \end{matrix} \right] ,
\end{equation}
where the electron (hole) Hamiltonian of the TI surface states is given by 
\begin{equation}
    \hat{h}_\mathrm{e / h}^\mathrm{TI} = \pm \hbar v_\mathrm{F} \big( - k_x \hat{\sigma}_y + k_y \hat{\sigma}_x \big) \mp \mu \hat{\sigma}_0 \pm U(x) - g \mu_\mathrm{B} B_y \hat{\sigma}_y / 2  . 
\end{equation}
Thereby, $ v_\mathrm{F} $ indicates the Fermi velocity ($ k_\mathrm{F} $ is the Fermi wave vector), $ \mu $ is the chemical potential, $ \hat{\sigma}_i $~($ \hat{\sigma}_0 $) corresponds to the $ i $th Pauli ($ 2 \times 2 $ identity) matrix, and $ U(x) = U_0 \hat{\sigma}_0 [\delta(x) + \delta(x - l_\mathrm{w})] $ introduces similar, deltalike barriers at the two interfaces between the weak link and the superconducting leads to model the reduced transparency of the weak link suggested by the measurements. 
Note that we describe the TI surface states in our model with a vector-product-like dispersion ($ \propto \vec{k} \times \hat{\vec{\sigma}} $), as its scalar-product-like counterpart ($ \propto \vec{k} \times \hat{\vec{\sigma}} $) will not produce anomalous $ \varphi_0 $-shifts for magnetic fields $ B_y \neq 0 $ along the transverse (in-plane) $ \hat{y} $-direction perpendicular to the current direction. 
The coupling of $ B_y $ with the TI surface states is included through a scalar Zeeman term~($ \mu_\mathrm{B} $ is the Bohr magneton); orbital components (i.e., Peierl's substitutions), as analyzed in Ref.~\cite{HimmlerPRR2023}, do not play a role for magnetic fields applied along $ y $ (i.e., in the plane of the unfolded wire); they only need to be included for magnetic-field components parallel to the current direction that will pierce the cross-section of the wire. 
The $ g $-factor of HgTe nanowires has been reported to vary significantly under modest gating, e.g., from zero to values above 600~\cite{Reuther2013}; for our simulations, we set $ g = 50 $, which is still rather moderate and could provide one explanation that our theory quantitatively underestimates the experimentally found $ |\varphi_0| $. 

The $ s $-wave superconducting pairing potential in the leads is given by 
\begin{equation}
    \Delta(x) = \Delta^{\ast} \big[ \Theta(-x) + \mathrm{e}^{\mathrm{i} \varphi} \Theta(x - l_\mathrm{w}) \big] 
\end{equation}
with the proximity-induced superconducting gap $ \Delta^{\ast} = 0.3 \, \mathrm{meV} $ and the phase difference $ \varphi $.

The numerical calculations shown in Fig.~3\textbf{e} were performed for a junction with the geometrical dimensions $ l_\mathrm{w} = 250 \, \mathrm{nm} $,  $ w_\mathrm{w} = 600 \, \mathrm{nm} $, and $ h = 80 \, \mathrm{nm} $. For the chemical potential and the SN barrier parameter, we chose $ \mu = 11.5 \, \mathrm{meV} $ and $ U_0 \approx 550 \, \mathrm{meV} \, \mathrm{nm} $, respectively, while $ \hbar v_\mathrm{F} = 330 \, \mathrm{meV} \, \mathrm{nm} $ corresponds to the Fermi velocity $ v_\mathrm{F} = 5 \times 10^5 \, \mathrm{m}/\mathrm{s} $.

To compare the TI with the parabolic-dispersion case (e.g., in the 2DEG of InAs quantum wells~\cite{Baumgartner2022,Costa2023}), we repeated the simulations for the same set of parameters replacing $ \hat{h}_\mathrm{e / h}^\mathrm{TI} $ by 
\begin{equation}
    \hat{h}_\mathrm{e / h}^\mathrm{parabolic} = \pm \hbar v_\mathrm{F} / k_\mathrm{F} ( k_x^2 + k_y^2 ) \hat{\sigma}_0 \mp \mu \hat{\sigma}_0 \pm U(x) - g \mu_\mathrm{B} B_y \hat{\sigma}_y / 2 + \alpha (- k_x \hat{\sigma}_y + k_y \hat{\sigma}_x ) ;
\end{equation}
the comparison of the resulting $ \varphi_0 $ with the TI case is also shown in~Fig.~3\textbf{e}. Note that the parabolic dispersion will only produce $ \varphi_0 $-characteristics in the additional presence of Rashba SOC with strength $ \alpha $~(we set $ \alpha = 10 \, \mathrm{meV} \, \mathrm{nm} $ as a realistic value~\cite{Baumgartner2022,Costa2023} in our simulations), and that we used the same Fermi energy $ \varepsilon_\mathrm{F} = \mu (T=0) = \hbar v_\mathrm{F} k_\mathrm{F} $ in the Dirac and parabolic cases to compare the resulting $ \varphi_0 $-shifts in magnitude.

Since TI surface states have a single Fermi surface with spin-momentum locking, they display a giant $\varphi_0/B_\text{ip}$ compared to the parabolic dispersion 2DEGs with Rashba, where the two Fermi surfaces with opposite spin-to-momentum orientation tend to mutually compensate magnetochiral effects. In the former case, $\partial \varphi_0 /\partial B_y$ is about $30$
times larger than in the latter. Moreover, the addition of an \textit{explicit} Rashba term in the Hamiltonian is nearly irrelevant in the TI case, while it is a necessary ingredient to observe magnetochiral effects in the parabolic case. On the other hand, the computed $\partial \varphi_0 /\partial B_y$ is still 45 times smaller than the measured one (notice the different abscissa scale in Figs.~3\textbf{a} and \textbf{e}).

\section{Spin texture of  2D surface states in (013)-grown H${\bf g}$T${\bf e}$}

The symmetry of the (013) HgTe surface is described by the point group $C_1$. It contains no spatial symmetry operations except for identity. This conclusion is  related to the size-quantization of surface carriers similar to (013) HgTe quantum wells where 
the point symmetry is also $C_1$, see Ref.~\cite{Ganichev2014} for a review.
As a result, the $\vec k$-linear Hamiltonian has the most generic form
\begin{equation}
\label{eq:h_generic_SM}
\hat{h}_\mathrm{e / h}^\mathrm{TI, lin} = \pm \hbar \sum_{\nu=x,y,z}\sum_{i=x,y} v_{\nu i} \hat\sigma_\nu k_i
\end{equation}
with six linearly-independent coefficients $v_{\nu i}$. 
Here the axes are chosen as $z \parallel [013]$, $x \parallel [100]$, and $y \parallel [03\bar{1}]$, i.e. $z$ is a normal to the surface, and 2D carriers propagate in the $(xy)$ plane.

It is convenient to rewrite this Hamiltonian using the combinations $(\hat\sigma_x k_y \pm \hat\sigma_y k_x)$, $(\hat\sigma_x k_x \pm \hat\sigma_y k_y)$ and $\hat\sigma_z k_{x,y}$:
\begin{equation}
\label{eq:h_generic_SM_1}
\hat{h}_\mathrm{e}^\mathrm{TI, lin} = \hbar v_{\rm R}(\hat\sigma_x k_y - \hat\sigma_y k_x) +\hbar v' (\hat\sigma_x k_y + \hat\sigma_y k_x)
+ \hbar \tilde{v} (\hat\sigma_x k_x + \hat\sigma_y k_y) + \hbar v_{\rm D}(\hat\sigma_x k_x - \hat\sigma_y k_y) 
+ \hat\sigma_z \hbar (v_{zx} k_x + v_{zy} k_y),
\end{equation}
where $v_{\rm R}, v', \tilde{v}, v_{\rm D}, v_{zx}, v_{zy}$ are also linearly-independent.
Here the contribution with $v_{\rm R}$ is the Rashba term, one with $v_{\rm D}$ can be called the Dresselhaus term,  $\tilde{v}$ gives the 2D Weyl (or chiral) term, and $v_{zx}$, $v_{zy}$ describe a coupling of the momentum with out-of-plane spin.

The microscopic origin of different terms in the spin-orbit coupling~\eqref{eq:h_generic_SM_1} is caused by structure, bulk and interface inversion asymmetries of the surface (SIA, BIA and IIA)~\cite{Ganichev2014}. BIA and IIA can give contributions to all six terms. Averaging the bulk Dresselhaus term caused by BIA over size-quantized motion of 2D carriers gives the terms $\propto v_{\rm D}$ and $\propto v_{zy}$ with $v_{zy}=3v_{\rm D}/8$.
SIA results in a contribution to the Rashba term. There is also a contribution arising from a joint action of SIA and cubic symmetry of the HgTe bulk crystal~\cite{Budkin2022}. Similar to (013) HgTe quantum wells, it results in the term $\propto v'$.

The spin texture of the surface states is determined by ratios of the quantities $v_{\nu i}$ in the Hamiltonian~\eqref{eq:h_generic_SM}, or, equivalently, of the terms in Eq.~\eqref{eq:h_generic_SM_1}. If the Rashba term is the most important, then the surface states have spin texture in the $(xy)$ plane perpendicular to their momenta (the angle between spin and momentum $\theta=90^\circ$). The other terms in the Hamiltonian change the spin texture.
If, for example, we take into account the Rashba and the 2D Weyl terms,  then 

\begin{equation}
\hat{h}_\mathrm{e}^\mathrm{TI, lin} = \hbar v_{\rm R}(\hat\sigma_x k_y - \hat\sigma_y k_x) 
+ \hbar \tilde{v}  (\hat\sigma_x k_x + \hat\sigma_y k_y) 
\\ \equiv \hbar V(\hat\sigma_1 k_y - \hat\sigma_2 k_x).
\end{equation}
Here the coordinates in the spin space read
\begin{equation}
\sigma_1 = \hat\sigma_x \cos\psi + \hat\sigma_y \sin\psi,
\quad \hat\sigma_2 = -\hat\sigma_x \sin\psi + \hat\sigma_y \cos\psi,
\end{equation}
with $\psi = \arctan(\tilde{v}/v_{\rm R})$, and the velocity $V=\sqrt{v_{\rm R}^2 + \tilde{v}^2}$.
This Hamiltonian describes 2D carriers with linear and isotropic dispersion.
The spin texture of the eigenstates lies in the surface plane at an angle $\theta=90^\circ - \psi$ to the wavevector $\vec k$.

\section{Wafer growth and sample fabrication}
\subsection{Wafer growth and band diagram}
\begin{figure}
    \centering
    \includegraphics[width=0.5\linewidth]{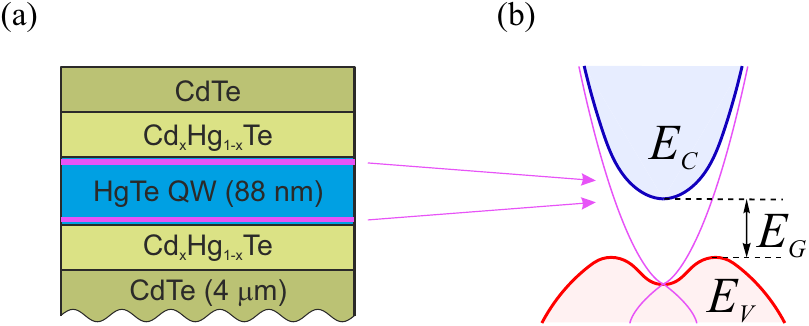}
    \caption{ (a) - Heterostructure layers grown on (013)-GaAs and 4~$\mu$m CdTe containing the 88-nm-thick strained HgTe film, barrier layers of 30-nm-thick Cd$_{\rm 0.65}$Hg$_{\rm 0.35}$Te below and 30-nm-thick Cd$_{\rm 0.65}$Hg$_{\rm 0.35}$Te above HgTe and the 40~nm thick CdTe cap layer. The topological surface states in HgTe are shown by the magenta lines.
    (b) Simplified electronic band structure of tensile-strained HgTe around the $\Gamma$-point as a function of an in-plane wave vector. The conduction band with bottom $E_c$ is shown in blue, the valence band with top $E_v$ in red and the topological surface states in magenta. The bulk gap $E_g$, opened by strain, is of order 15~meV. The Fermi level position is expected to lie in the valence band near $E_v$.}
    \label{fig:HgTe-Crosssection}
\end{figure}

The HgTe strained films were grown by molecular beam epitaxy on a (013)-oriented GaAs/CdTe substrate. A vicinal surface orientation was chosen in order to reduce the formation of surface defects that typically emerge during the growth process \cite{Kvon2009}. The layer structure is shown in Fig.~\ref{fig:HgTe-Crosssection}(a). The central part of the heterostructure is an 88~nm HgTe film encapsulated between 30~nm Cd$_{\rm 0.65}$Hg$_{\rm 0.35}$Te barrier layers, which at the same time act as buffer layers, smoothing the transition from HgTe to CdTe. The structure is capped with a 40~nm CdTe layer. 

The surface states are located near the interfaces between the HgTe and CdHgTe barriers as well as on the etched mesa side facets, directly connecting the top and bottom surface states. The simplified band diagram is shown in Fig.~\ref{fig:HgTe-Crosssection}(b).
Since the lattice constants of HgTe and the underlying CdTe layer are slightly different (a$_{\rm CdTe}$ = 0.648~nm and a$_{\rm HgTe}$ = 0.648~nm), the typically semimetallic HgTe layer is under tensile strain of about 0.3\%, which opens a gap between the valence and conduction bands \cite{Bruene2011, crauste2013topologicalsurfacestatesstrained, WuEPL2014, Savchenko2019}. 
As the structure is undoped, we assume that the Fermi level is located in the vicinity of the valence-band maximum. The transport properties are discussed further.

\subsection{Sample fabrication}
The following steps were performed to build the SQUID device.

\subsubsection{Defining the HgTe nanowire}
\begin{enumerate}
    \item The starting point is a chip with stack sequence (top to bottom) 40~nm CdTe, 30~nm Cd$_{0.65}$Hg$_{0.35}$Te, 88~nm HgTe, 20~nm Cd$_{0.65}$Hg$_{0.35}$Te, 4~$\mu$m CdTe, GaAs substrate.
    \item Using electron beam lithography (EBL), the negative mask for the nanowire is patterned. 
    The nanowire width is $w\ti{w}=715\,$nm for sample A and $w\ti{w}=1.1$\,µm for sample B.
    \item Applying a bromine etching solution removes the layer stack down to the 4µm CdTe all over the chip except for the nanowire structure
    \item After rinsing off the resist, the structure is ready for the next EBL step.
\end{enumerate}

\subsubsection{Contacting the HgTe nanowire with Nb leads}

\begin{figure}
    \centering
    \includegraphics[width=0.99\linewidth]{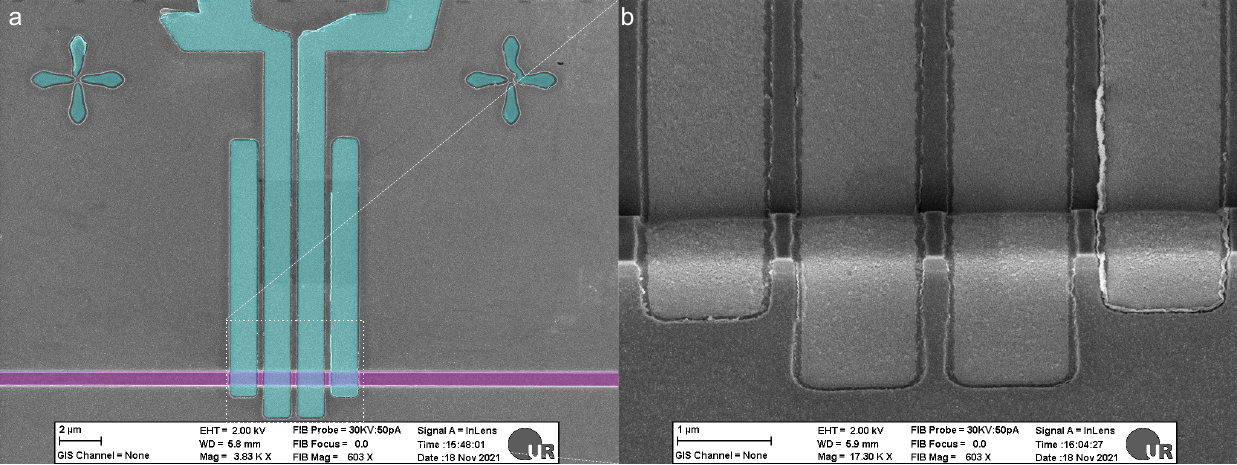}
    \caption{(a) SEM image showing sample A after the Nb contact lead lift-off. False color highlights Nb in cyan, and the remaining HgTe layer forming the wire in purple. (b) Zoom SEM image of the contact area, taken at a tilted angle.}
    \label{fig:Nb-Evap}
\end{figure}

\begin{enumerate}
    \item Via EBL, the contact electrodes for the Nb leads are written onto the wire
    \item Bromine wet etching removes the CdTe and 30~nm Cd$_{0.65}$Hg$_{0.35}$Te layers completely and the HgTe layer partially 
    \item The sample is immediately placed in the vacuum of the load lock of the evaporation chamber
    \item Argon milling is applied to remove oxides from the HgTe contact area. For this a Kaufman source is used to minimize the impact on the material.
    \item Metallization consists of 2-3nm of Ti as an adhesive layer, followed by 80nm of Nb, with a 3nm Pt capping layer on top.
    \item After performing an ultrasonic-assisted lift-off, the resulting structure corresponds to fig. \ref{fig:Nb-Evap}
\end{enumerate}

\subsubsection{Defining the SQUID loop and reference junction}

\begin{figure}
    \centering
    \includegraphics[width=0.99\linewidth]{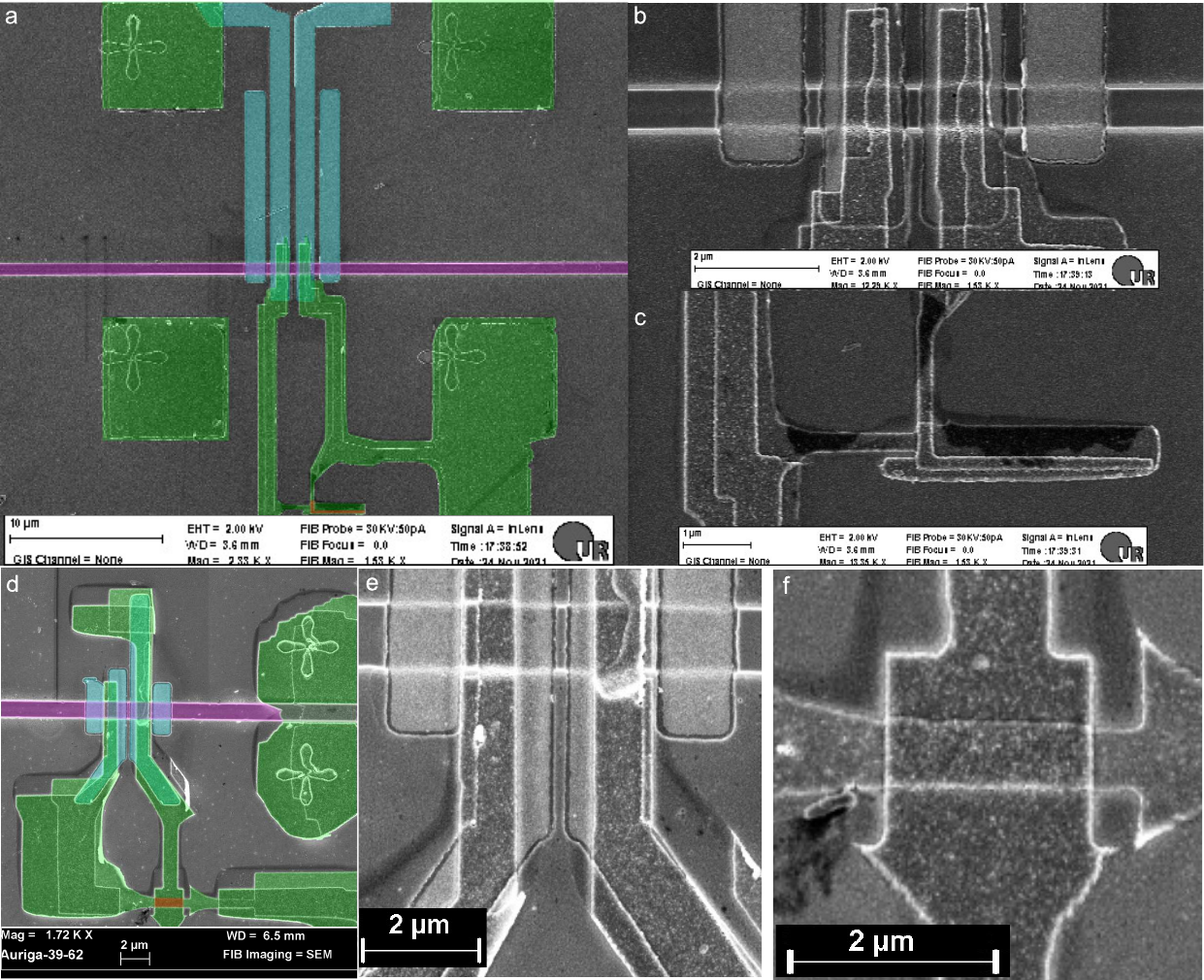}
    \caption{SEM images showing sample A (a,b,c) and sample B (d,e,f) after the Al lift-off. A large overview is presented in panels a and d, where the false color cyan indicates Nb, purple indicates the remaining HgTe layer stack, green indicates Al, and orange indicates the area of the Al/AlOx/Al reference junction. A zoom of the wire areas is shown in panels b and e. The reference junctions are shown magnified in panels c and f.}
    \label{fig:FinishedSample}
\end{figure}

\begin{enumerate}
    \item For this lithography, a resist several µm thick is necessary to allow the Manhattan shadow evaporation.
    \item EBL defines the reference junction as a crossover area of two perpendicular trench lines. In the same step, the remaining part of the SQUID loop is also written. For the SQUID leads, sufficient width is crucial to avoid gaps caused by shadow evaporation.
    \item In the load lock of the evaporation chamber, Kaufman Argon milling is applied to clean the surface.
    \item The first Al layer is evaporated using electron beam evaporation onto the cooled substrate ($\sim$-90°C). Here, the evaporation angle is tilted (by 27° for sample A and 49.5° sample B) relative to the plane normal towards one of the lines that cross at the reference junction.
    \item A controlled in situ oxidation step is done in the load lock to create the barrier oxide for the Al/AlOx/Al reference junction
    \item The second Al layer is evaporated onto the cooled substrate ($\sim$-90°C). The evaporation angle is tilted (by 27° for sample A and 32.4° sample B) relative to the plane perpendicular to the tilt direction of the first evaporation step.
    \item After lift-off, the structure is complete and ready to be bonded and cooled down. A SEM picture is shown in Fig. \ref{fig:FinishedSample}.
\end{enumerate}

\subsubsection{Defining the on-chip lines and packaging}
To minimize screening currents across the chip and improve B-field 
homogeneity, 
the use of superconducting material in the surroundings of the device was avoided as much as possible. In fact, field expulsion due to Meissner effect modifies the field lines,  making it more difficult to achieve a reliable compensation of the out-of-plane field, see Section \ref{sec:compen}. Accordingly, beyond a distance of 100~µm from the SQUID, the Al leads were continued as Au leads. 
For the same reason, the bonding wires were also made of gold.

\section{Basic characterization of the quantum well and the superconductor}

\subsection{Quantum well}

\begin{figure}
    \centering
    \includegraphics[width=0.3\linewidth]{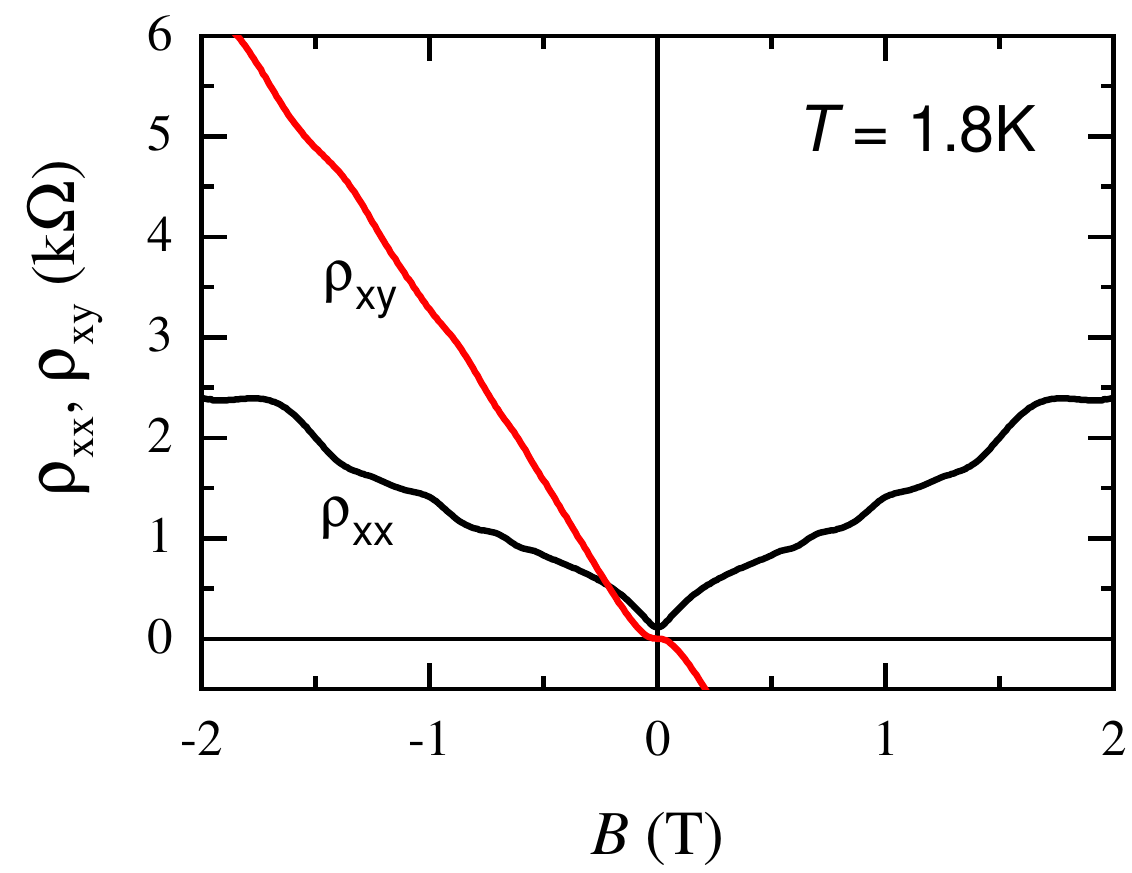}
    \caption{The transport response of a macroscopic HgTe Hall bar in classical magnetic fields. Due to the presence of two types of carriers (surface electrons and bulk holes), a nonlinear Hall effect ($\rho_{xy}$, shown in red) and a pronounced positive magnetoresistance ($\rho_{xx}$, shown in black). Fitting the measured dependencies using a two-component Drude model (not-shown) allowed us to extract the electron and hole densities and mobilities, which amount to $N_s = 0.8 \cdot 10^{11}~\text{cm}^{-2}$, $P_s = 2.5 \cdot 10^{11}~\text{cm}^{-2}$, $\mu_e = 280 \cdot 10^{3}$~cm$^2/$V$\cdot$s, $\mu_p = 90 \cdot 10^{3}$~cm$^2/$V$\cdot$s.
    }
    \label{fig:DrudeTransport}
\end{figure}

The quantum well under study belongs to the class of three-dimensional topological insulators and is characterized by the dispersion relation shown in Fig.~\ref{fig:HgTe-Crosssection}(b). The system contains a valence band and a conduction band separated by an indirect bulk gap of approximately 15~meV, as well as surface states with a quasi-Dirac dispersion (shown in magenta in Fig.~\ref{fig:HgTe-Crosssection}(b)). The transport properties of thin HgTe films have been extensively investigated in our previous works \cite{Kozlov2014, Kozlov2016, Ziegler2020, Hartl2025}. These studies demonstrated that the surface states exhibit high mobility (up to 600$\cdot$10$^{3}$~cm$^{2}$/V$\cdot$s).
In ungated samples \cite{Kozlov2014}, the Fermi level lies close to the top of the valence band; under these conditions, surface electrons may coexist with bulk holes, although the contribution of the surface electrons to the conductivity remains dominant, which has also been demonstrated in studies of SNS-junctions \cite{HimmlerPRR2023}. The coexistence of electrons and holes manifests itself in the dependencies $\rho_{xx}(B)$ and $\rho_{xy}(B)$ shown in the fig.~\ref{fig:DrudeTransport}. Based on fitting the experimental curves with a two-component Drude model (following Ref.~\cite{Kozlov2014}), we obtained a net (i.e. both top and bottom) surface electron density of $N_s = 0.8 \cdot 10^{11}~\text{cm}^{-2}$ and average mobility of $\mu = 280 \cdot 10^{3}$~cm$^2/$V$\cdot$s. We note that in the devices under study, the carrier properties may differ somewhat due to the effects of nanostructuring, which can lead to charge accumulation as well as to a redistribution of mechanical strain, both of which may influence the electronic spectrum.

\subsection{Aluminum characterization\label{app:alcharact}}
\begin{figure}
    \centering
    \includegraphics[width=0.6\linewidth]{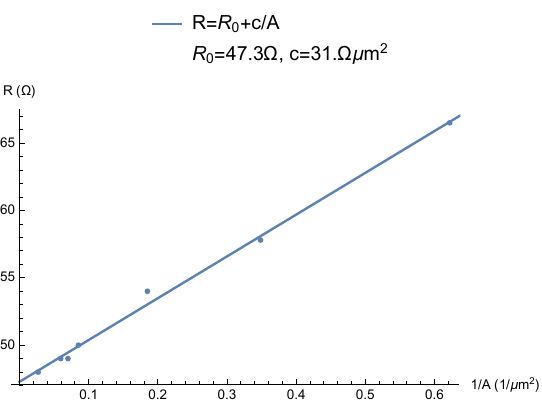}
    \caption{Resistances of single SIS Aluminum junctions with different junction area size $A$ at room temperature. The constant offset of $47\,\Omega$ is ascribed to the aluminum stripes connecting the top and bottom side of the junction.}
    \label{fig:linresArea}
\end{figure}

To  characterize the superconducting properties of our reference sample, we analyze the measurement data of sample T55.  This sample contains several single SIS junctions  with junction areas ranging from $1.61\,$µm$^2$ to $37.07\,$µm$^2$. Fig \ref{fig:linresArea} shows the total resistances of the junctions as a function of the inverse junction area. The behaviour is linear with an offset of $47\,\Omega$ that can be ascribed to the resistance of the single-layer Al stripes leading to each layer of the junction. The single-layer Al leads have 10 squares each, with the first layer having a thickness of $31\,$nm and the second layer $67\,$nm, resulting in $\rho\ti{n}=4.0\,$µ$\Omega$cm.

\begin{figure}
    \centering
    \includegraphics[width=0.6\linewidth]{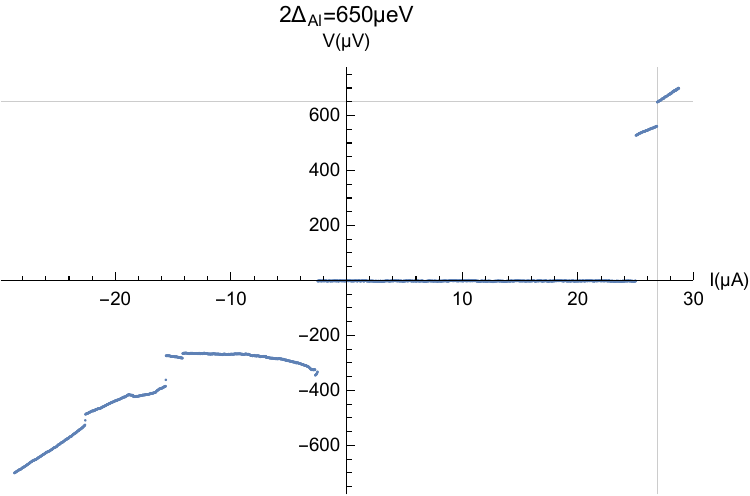}
    \caption{IV of single SIS Aluminum junction at $\sim 50\,$mK for sample T55.}
     \label{fig:getDelta}
\end{figure}

Besides the normal conductivity, the superconducting gap $\Delta\ti{Al}$ is of great importance. 
We derive it from half the voltage at which a SIS junctions shows ohmic behavior. A measurement from sample T55 yields $\Delta\ti{Al}=325\,$µeV see fig. \ref{fig:getDelta}.

\section{Measurement techniques and data evaluation}

\subsection{Compensating out of plane magnetic field component and flux removing procedure}
\label{sec:compen}
Probing the $B_\parallel$-dependent $\varphi_0$-shift in an asymmetric DC SQUID setup requires precise control of the out-of-plane field component ($B_z$) due to unavoidable imperfect alignment of the $B_\parallel$ field with respect to the sample surface. Thus, it is crucial to develop a protocol that makes it possible to null $B_z$ with an accuracy much better than a quantum of flux over the entire SQUID loop, which corresponds to $\sim30\,$µT.

\begin{figure}
    \centering
    \includegraphics[width=0.9\linewidth]{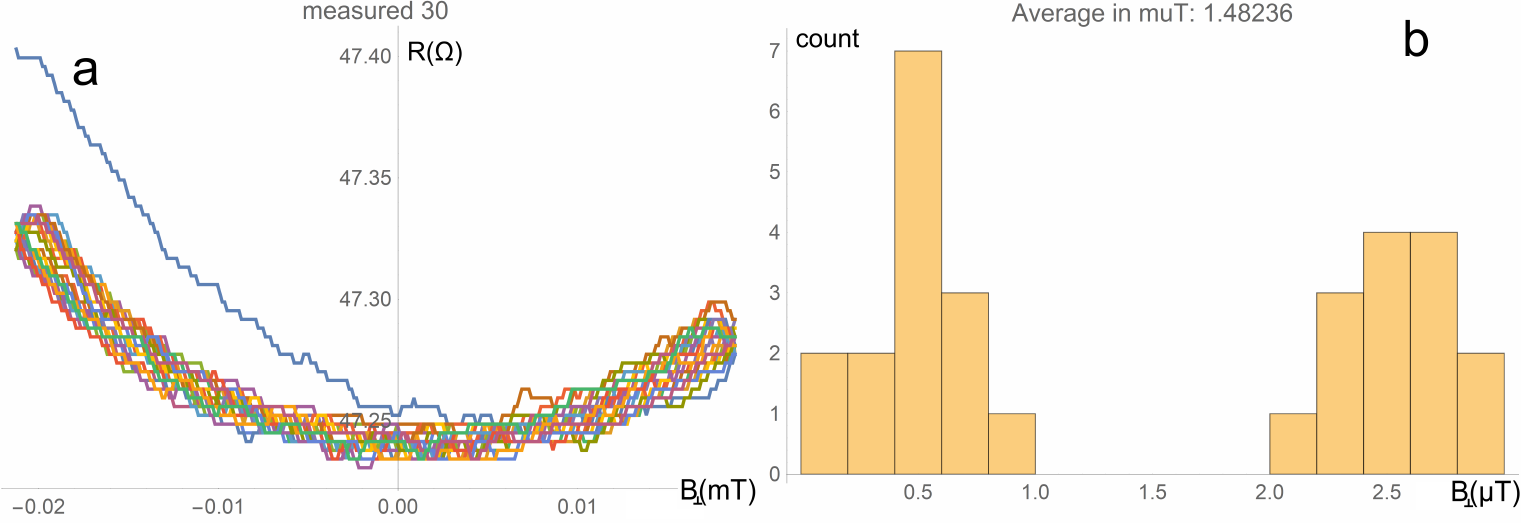}
    \caption{(a) Thirty $R(B_{\perp})$ curves measured in step 6 of Method 1, see text. Measurement data of step 6 $R(B_\perp)$. The scale of $B_\perp$ is the original value taken from the instruments output. (b) Histogram of the minima positions for parabolic fits of the data of panel a. The outlier curve (blue) is left out. The switching distribution is bimodal: the left (right) peak in the histogram corresponds to a leftwards (rightwards) sweep. This separation between the peaks proportional to the sweep speed, which lies at $\approx1\,$µT/s.}
    \label{fig:finecomp}
\end{figure}

To this end, we make use of the sharp dependence of the Al resistance in the fluctuation regime at fields and temperatures close to the phase boundary. In particular, for temperatures close to $T\ti{c}$, the resistance has a parabolic dependence on $B_z$, whose minimum allows us to determine the effective $B_z=0$, see Fig \ref{fig:finecomp}. In particular, for temperatures such that $R=0.8R_n$, the curvature of the parabolas is sufficiently sharp to allow us to determine, by fitting, the minimum with an accuracy better then $0.5\,$µT. 

In detail, the compensation protocol we used is the following:
\begin{enumerate}
    \item The  in-plane magnetic field $B_\parallel$ is set to the desired value.
    \item The cold finger of the dilution refrigerator is heated to $200-300\,$mK.
    \item The chip-carrier is heated until the resistance of the total SQUID, dominated by the resistance of the reference junction is at about $80\%$ of its $R\ti{n}$. An additional resistor of $~1\,$k$\Omega$, mounted directly at the chip carrier, is used for heating.
    \item To obtain a first approximation of the compensation field, $B_\perp$ is measured over a range of $\pm300\,$µT using both forward and backward sweeps. Due to hysteresis in the superconducting coil, the minima of the two sweeps do not coincide. The initial working point for $B_\perp$ is then set to the average of the minima from the forward and backward sweeps.
    This step is essential because it establishes a starting point for the subsequent fine adjustment. Without it, the compensation field determined later may deviate significantly from the correct zero-field condition, making it difficult to maintain the target $R=0.8R\ti{n}$
    \item Once the initial $B_\perp$ working point has been established, the heating power applied to the chip carrier resistor is adjusted so that the SQUID resistance returns to approximately $R(T)= 0.8R\ti{n}$. It is important to allow sufficient time for the sample temperature to stabilize and for the system to reach thermal equilibrium before proceeding to the next step. This ensures that the resistance minimum corresponds accurately to the intended working point.
    \item To determine the compensation field with high precision, $R(B_\perp)$ is measured in a narrow range of $±20\,$µT around the preliminary working point. A total of 30 sweeps are performed, consisting of 15 forward sweeps and 15 backward sweeps. Sweeps are conducted rapidly (less than $50\,$s each) to minimize the influence of temperature drift. Occasionally, individual traces are distorted by transient temperature fluctuations; these outliers, such as the blue curve in Fig. \ref{fig:finecomp}a, are identified and excluded from the analysis. Each remaining trace is fit with a parabola to extract the field corresponding to the resistance minimum. The distribution of minima from forward and backward sweeps is shown in the histogram in Fig. \ref{fig:finecomp}b, forming two separate clusters separated by approximately $2\,$µT due to hysteresis in the superconducting coil. The compensation field is then defined as the midpoint between these two clusters, providing the most accurate estimate of $B_\perp=0$.
    \item After the initial calibration of $B_\perp$, the resistor on the chip carrier is used to heat the sample well above the aluminum Tc, allowing any trapped vortices to escape. Performing this heating after the initial sweeps ensures that the preliminary zero-field working point is first identified using the SQUID’s resistance response. The system is then left to thermally stabilize. Higher heating pulses, sufficient to also exceed the Nb 
Tc, were tested, but no significant changes were observed aside from longer cooldown times.
    \item Steps 5–7 are repeated up to three times. The first iteration may be affected by residual trapped vortices that are out of equilibrium. After heating, these vortices vanish, so the second iteration begins from a condition free of trapped vortices, producing more reliable results. The third and final round serves as a check for the convergence and stability of the compensation procedure.
    \item As a final safeguard, an even stronger heating pulse is applied at the end of the last iteration of Step 6 to ensure that no residual vortices remain. After this pulse, the system is allowed to thermally stabilize and then cooled back down to the measurement temperature.  When the in-plane field $B_{||}$ is changed, the zeroing procedure must be repeated to recalibrate $B_\perp=0$.
\end{enumerate}

\begin{figure}
    \centering
    \includegraphics[width=0.6\linewidth]{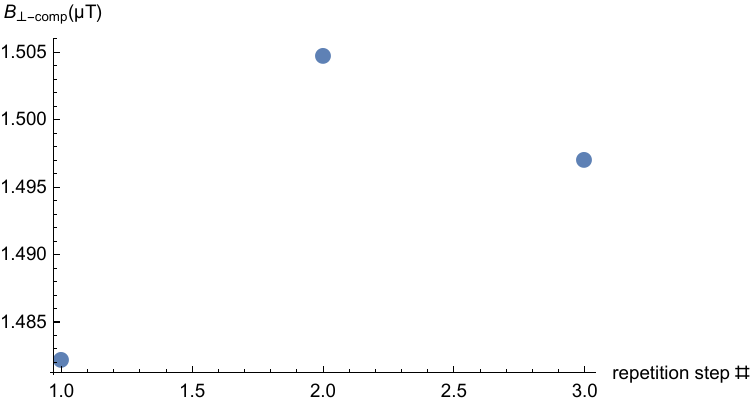}
    \caption{Compensation values obtained in step 7 of Method 1,  for the first, second, and third repetition.}
    \label{fig:compvalues}
\end{figure}

The compensation procedure described above provides a reliable and verifiable method to minimize the influence of trapped vortices on the determination of $B_\perp=0$. Figure \ref{fig:compvalues} illustrates the reproducibility of the procedure by showing the results for three consecutive sequences of steps 5–7. From these data, we estimate that the reproducibility - i.e., the precision with which the compensation field can be reproduced - is approximately $20\,$nT. As expected, the difference between the second and third iteration is smaller than that between the first and second iteration, reflecting the progressive removal of trapped vortices. Importantly, all differences are on the order of a few tens of nT, which is much smaller than the overall precision of the procedure.

It is important to distinguish precision from accuracy. Precision refers to the width of the distribution of compensation values obtained from repeated routines, whereas accuracy denotes the deviation of the mean value from the true zero-flux condition. Accuracy can be affected by factors such as inhomogeneities in the applied compensation field (for example, slight variations of the perpendicular component of $B_{\parallel}$ or the compensation field across the sample surface), or other experimental imperfections. To independently verify the accuracy of the compensation procedure, we implemented additional methods based on devices located at positions far from the device under study; these alternative procedures are described in the next section.


\subsection{Crosscheck of different compensation methods}
\label{app:crosscheckMethods}
In what follows, we shall indicate the main compensation method (described in the previous section) as \textbf{Method 1}. The alternative methods we adopted are described in the following. These are used to validate the main Method 1 and provide an estimate of its accuracy.

\begin{figure}
    \centering
    \includegraphics[width=0.95\linewidth]{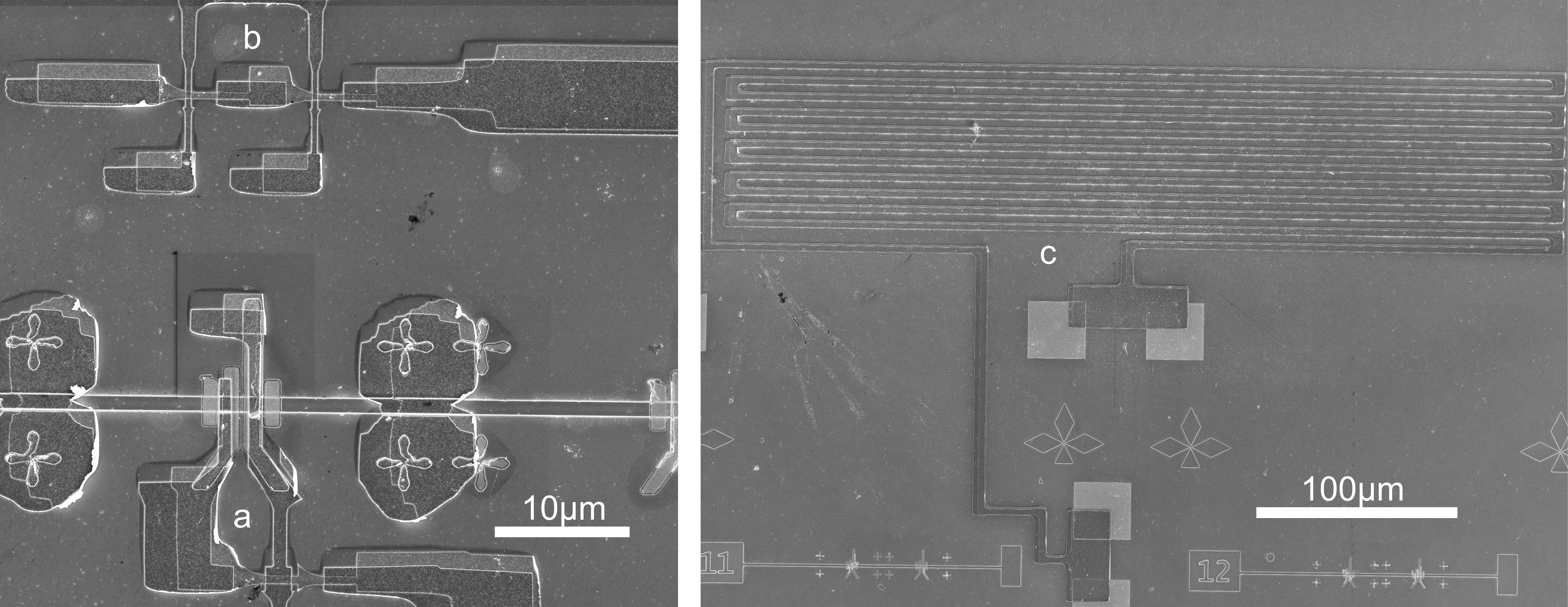}
    \caption{(Left) SEM picture of the area near the device under study (sample B) labelled as $\mathsf{a}$. On the top part of the SEM scan, the symmetric SIS SQUID used in Method 3 is visible. (Right) SEM picture of the meander device in sample B used for the compensation Method 2 (see text).}
    \label{fig:compensation-3geometries}
\end{figure}

\textbf{Method 2.} A separate aluminum meander structure located $240\,$µm away from the HgTe SQUID under study (see Fig.\ref{fig:compensation-3geometries}c) is used. The perpendicular field 
$B_\perp$ is adjusted to minimize the resistance of this meander near its aluminum Tc. This provides an independent measure of 
$B_\perp=0$.

\textbf{Method 3.} We use a symmetric SQUID consisting of two identical junctions similar to the reference junction of the device under study and located at a $35\,$µm distance from it. This SQUID is used as a magnetometer: its total critical current is probed parallel to that of the HgTe SQUID. The emerging pattern of the two identical SIS junctions is then analyzed to obtain $B_\perp=0$. This Method is fundamentally different to the others since it operates in the fully superconducting state.

\textbf{Method 4.} It works the same way as \textbf{Method 1}, but instead of probing $R(B_\perp)$ for the HgTe SQUID,  the SIS SQUID mentioned in \textbf{Method 3.} is measured. 

\begin{figure}
    \centering
    \includegraphics[width=0.95\linewidth]{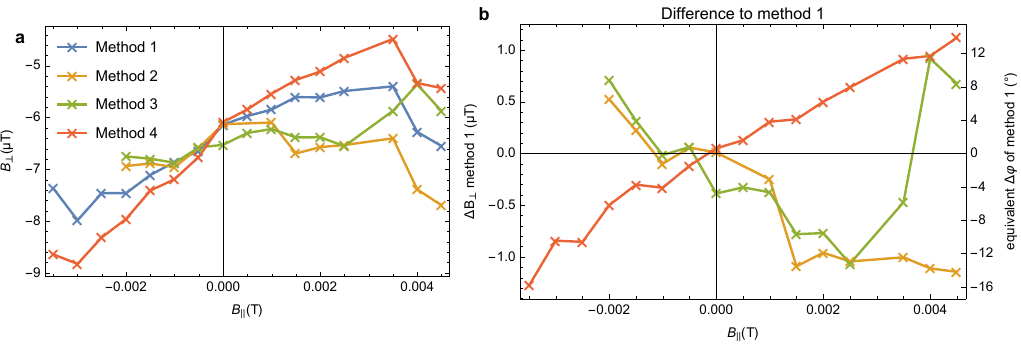}
    \caption{Comparison of different compensation methods performed on sample B. (a) The graph shows the $B_{\perp}(B_\parallel)$ curves for the four different compensation geometries, where $B_{\perp}$ is the applied perpendicular field that produces full compensation according to each method. (b) The graph shows the difference with respect to the results of Method 1, plotted for the other Methods. To express the magnetic field in terms of phase (right-hand axis), the periodicity of $~29\,$ µT of the HgTe SQUID was used.}
    \label{fig:compensation-crosscheck}
\end{figure}

A comparison of $B_{\perp}(B_\parallel)$ for the different Methods is shown in Fig.~\ref{fig:compensation-crosscheck}.
From the graph we can conclude the following. (i) The scatter of the data points is always better than 500 nT. (ii) The deviations with respect to Method 1 (shown in panel b) are roughly proportional to the in-plane field itself, which may indicate field inhomogeneity. However, the difference between Method 2 and 3 is very small even though they are based on devices which are very far apart. On the other hand, there is a large difference between Method 3 and 4, which are based on the same device (Al-SIS SQUID). 

From the comparison of the different methods, we can deduce that for an in-plane field up to 4~mT (which is roughly the field where the jump occurs in Fig.~3c, and of the same order as the full abscissa range of the Fig.~3a,c) different compensation methods, based on devices which are far apart from the HgTe SQUID, yield the same compensation field within at most 1~$\mu$T, corresponding to 2-3\% of the field producing a flux quantum in the HgTe SQUID. We take this 1~$\mu$T as the accuracy of the field compensation method.  The accuracy for Method 1 is presumably better than that, since it works directly on the device under study, so field inhomogeneity play a minor role. This means that the $\varphi_0$ shift observed in Fig.~3a of the main text, which is 0.8$\pi$ at 4~mT, cannot be due to a spurious out-of-plane component of the applied in-plane field. 



\subsection{Role of perpendicular critical field}
For both the preceding discussion of $B_\perp$ compensation and the subsequent analysis of $\varphi_0$-shift jumps, the perpendicular critical field of the aluminum film is crucial. Since we consider a thin film, we focus on the perpendicular case and verify that the critical field is well above the precision of our compensation method. This ensures that the Meissner effect can fully expel flux from the SQUID leads when the sample is cooled after the final compensation step.
We calculate key superconducting parameters using conservative approximations that tend to underestimate the critical field, providing a minimum estimate. Following \cite{Romijin1982}, the effective penetration depth is
\begin{align}
    \lambda\ti{eff}=\sqrt{0.18\frac{\rho\hbar}{k\ti{B}T\ti{c}\mu_0}}=109\,\text{nm},
\end{align}
where we use $\rho=4.0\,$µ$\Omega$cm and $T\ti{c}=1.2\,$K, although the measured superconducting gap would suggest a larger value.
The coherence length of the clean film is
\begin{align}
    \xi_0=\frac{\hbar v\ti{F}}{\pi\Delta}=1031\,\text{nm},
\end{align}
using $v\ti{F}=1.3\cdot10^6\,$m/s and $\Delta=325\,$µeV.
Following \cite{Romijin1982}, using the electronic mean free path
\begin{align}
    l\ti{el}=\rho^{-1}\cdot4\cdot10^{-16}\,\Omega\text{m}^2=10\,\text{nm},
\end{align}
we get the effective coherence length 
\begin{align}
    \xi=\sqrt{\xi_0l\ti{el}}=92\,\text{nm}.
\end{align}
Following \cite{Tinkham1983}, the perpendicular lower critical field of a thin aluminum film, above which it behaves as a type-II superconductor, is given by 
\begin{align}
    H\ti{c1}=\ln{\kappa}\frac{\Phi_0}{4\pi\lambda\ti{Pearl}^2},\label{eq:Hc1}
\end{align}
where the Ginzburg-Landau parameter is
\begin{align}
    \kappa=\frac{\lambda\ti{eff}}{\xi}=2.1,
\end{align}
and the Pearl penetration depth is
\begin{align}
    \lambda\ti{Pearl}=\frac{\lambda\ti{eff}^2}{t},
\end{align}
where $t$ is the thickness of the film. The thickness of the film is different for sample A and B, where the thinnest film, which lies at the edges of the double film structure, is given by
\begin{align}
    t\ti{min,A/B}=54/42\,\text{nm}.\label{eq:tAB}
\end{align}
Using this in eq. \ref{eq:Hc1} leads us to $B\ti{c1, A/B}=260/162\,$µT.

In the Meissner regime, screening currents concentrate magnetic flux near the edges of the film stripe. To account for this effect, we introduce a demagnetization factor 
$N$, yielding a reduced effective critical field, such that
\begin{align}
    B\ti{c1-eff}=(1-N)\cdot B\ti{c1}.
\end{align}
Modelling the stripe as a flat ellipsoid with dimensions $a>b\gg c$, \cite{Osborn.1945} the demagnetization factor is
\begin{align}
    N = 1 - \frac{cE}{a \left(1 - e^2\right)^{\frac{1}{2}}},
\end{align}
where E is the complete elliptic integral with the argument
\begin{align}
    e=\sqrt{1-b^2/a^2}.
\end{align}
For the thickness $c\ti{A/B}=133/110\,$nm we use the total thickness of the film in the center while $a\ti{A/B}=15/8.6\,$µm and $b\ti{A/B}=1.2/6.5\,$µm are estimated from the dimensions of the leads on the left hand side in Fig. \ref{fig:geometric-overview} a and b. This gives effective perpendicular critical fields of $B\ti{c1-eff,A}=29\,\text{µT}$; $B\ti{c1-eff,B}=3.5\,\text{µT}$, which are significantly larger than the precision of our field compensation ($\approx1$~\textmu T see \ref{app:crosscheckMethods}). These values should be regarded as lower bounds for the critical field.

\subsection{$\varphi_0$-jumps}
In the main text, we showed that the $\varphi_0$-shift features jump at certain well-defined values of the in-plane field. A $0-\pi$-transition is indeed expected as a consequence of the Zeeman field\cite{Yokoyama2014}, but such transition should be accompanied by a sharp, cusp-like minimum in the critical current (i.e., the amplitude of the CPR). As pointed out in the main text, the amplitude of the CPR, to very good approximation, is constant within the range of Fig.~3\textbf{a}. This is also what our tight-binding simulations predict, see Fig.~3f of the main text. 

\begin{figure}
    \centering
    \includegraphics[width=0.99\linewidth]{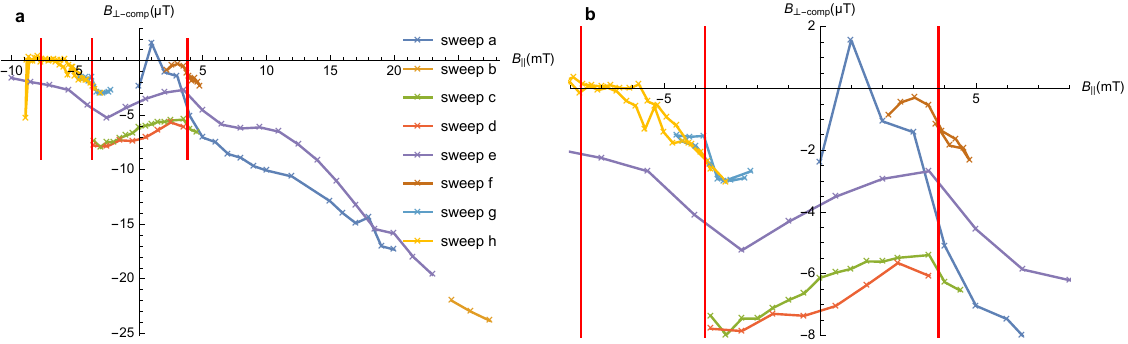}
    \caption{Multiple $B_{\perp-\text{comp}}$ sweeps were done at sample B, some of which were designed as back-and-forth sweeps (consecutive points are connected by a line). Panel b shows a zoom to the range that was pictured in Fig.~3a.
    \label{fig:compensation_jumps}}
\end{figure}

Therefore, we tried to identify extrinsic mechanisms that could produce such $\varphi_0$ jumps. First, we looked for possible discontinuities in the compensation field versus the applied in-plane field. In Fig.~\ref{fig:compensation_jumps} we show the compensation curves for many different sweeps. Notice that the ordinate scale is three orders of magnitude smaller than the abscissa scale. In the curves of Fig.~\ref{fig:compensation_jumps} we marked with red lines the in-plane field values where the $\varphi_0$ jumps occur, see Fig.~3c of the main text. We do not observe any discontinuity or abrupt behavior at these field. Thus, we can exclude that the jumps arise from a sudden change in the out of plane field (due, e.g., to a ferromagnetic switch of a nearby object, or to vortex entry of vortex redistribution within the bonding wires or to vortex entry within the superconducting coil of our cryostat). 

Another possibility is that our calibration Method 1, based on finding the resistance minimum of the Al branch of the SQUID might not identify exactly the field producing zero flux in the SQUID \textit{if the compensating field is not homogeneous in space}. However, the comparison with the other methods shown in Fig.~\ref{fig:compensation-crosscheck} indicates that the inhomogeneity cannot be much larger than a one or two microteslas over distances of hundreds of micrometers. On the other hand, we do observe differences in the compensation field between Method 3 and 4: hence a discrepancy between the magnetic field that minimizes the Al-branch resistance and that which produces exactly zero flux in the SQUID is conceivable. Such a discrepancy, however, cannot realistically exceed two microteslas.
Such field magnitude is far smaller than a flux quantum in the SQUID, and \textit{a fortiori} , far smaller than a flux quantum in the Nb leads. Thus, it is not realistic that the jumps
correspond to an Abrikosov vortex nucleation near the weak-link, which could alter the field profile deep in the superconducting regime. 

\begin{figure}
    \centering
    \includegraphics[width=0.99\linewidth]{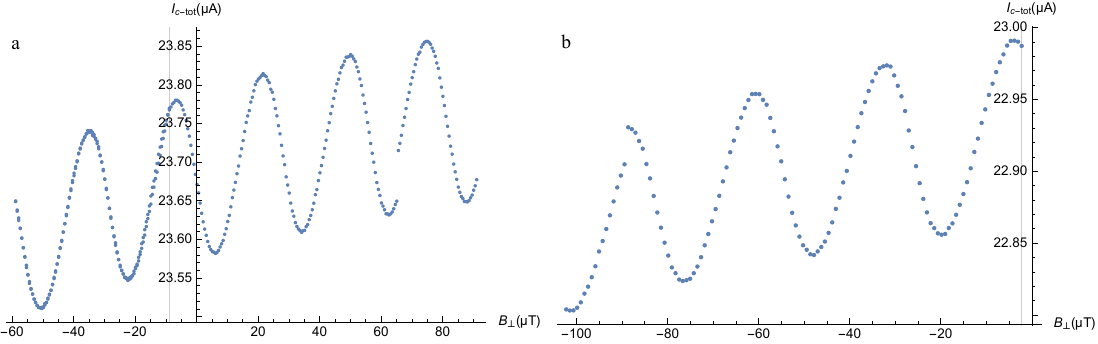}
    \caption{Example of raw CPR data both of sample B, without background subtraction, showing phase glitches. The compensation point in $B_\perp$ is indicated by a vertical line. 
    \label{fig:glitches}}
\end{figure}

Indeed, we do observe phase glitches when probing the CPR at magnetic fields away from the compensation point, see Fig.~\ref{fig:glitches}. These, however, occur randomly in both the magnetic field value at which CPR glitches appear and the phase value at which the CPR suddenly changes. The important things to keep in mind are: (i) glitches as those in Fig.~\ref{fig:glitches} are occurring randomly and are relatively rare: only a small fraction of the CPR traces is affected by a glitch; (ii) glitches are promptly identified, the portion of CPR trace after the glitch is ignored and does not play a role in the determination of $\varphi_0$; (iii) independently of the presence of a glitch, after a CPR trace a field compensation routine is applied which resets the phase of the CPR trace; (iv) as a consequence of the two previous points, the determination of $\varphi_0$ does not depend on the occurrence of such glitches: the absolute phase of a CPR trace can always be identified unequivocally and leads to a well defined value for $\varphi_0$.

Thus, the reproducible jumps observed in $\varphi_0(B_{ip})$ in Fig.~3c of the main text have nothing to do with the randomly occurring glitches in the CPR traces when sweeping $B_\perp$.

\subsection{Probing and analyzing $I\ti{c}(B_\perp)$}
Another challenge we faced is the hysteretic behavior of the SIS reference junction due to its underdamped nature. Random switching drives the junction into the normal conducting state. Because of hysteresis for currents $I\ti{retrap}<I<I\ti{c}$ the junction does not return to the superconducting state even though the applied current is below the critical current. 
As a result, increasing the integration time per data point does not help reduce data scatter: the longer an $IV$ trace takes, the farther the switching point drifts from the true $I\ti{c}$. To overcome the hysteresis problem, we switched from point-by-point IV measurements to a fast, continuous current sweep. We used a triangular-wave current generated by an arbitrary waveform generator (AWG). This approach repeatedly sweeps the current through the junction, allowing us to record many IV traces in rapid succession.

For sample B, we slightly modified the waveform: the slope of the triangle was made steeper for $|I| \lesssim 0.8I\ti{c}$. This ensures that more measurement points are collected close to the critical current, where the extraction of $I\ti{c}$ is most sensitive.

Using the AWG comes with one drawback: the voltmeter can no longer perform averaging before the data are transferred to the computer, so the raw voltage signal must be saved at high sampling rate and averaged only later during data processing. This results in much larger data files.

Transferring both voltage and current data would slow down the acquisition prohibitively—already the voltage data alone consumed about 30–40$\%$ of the total acquisition time. Therefore, we did not record the current trace directly. Instead, the current bias was defined reliably by choosing a sufficiently large series resistor, so that the current can be inferred unambiguously from the applied voltage.

The voltage response was recorded using a digital oscilloscope or by a Keysight  34470A (the latter with a sampling rate of 20000 data points per second). With this setup, we can typically acquire about 500 IV curves within approximately 40 seconds. These curves are then processed to extract the average critical current and the width of its statistical distribution.

\subsection{Evaluation of CPRs and $I\ti{c-ref}$ background removal}
In an ideal asymmetric SQUID used to probe a CPR, the measured critical current would be $I\ti{SQUID}(2\pi\cdot B_\perp/B\ti{period})=I\ti{c-ref}+\text{CPR}\ti{probe}(\varphi+\pi/2)$. In practice, however, the critical current of the reference junction depends of the flux biasing perpendicular magnetic field: $I\ti{c-ref}=I\ti{c-ref}(B_\perp)$. This field-dependent background must be removed in order to isolate the CPR of the probe junction.

We perform the background subtraction as follows:
\begin{enumerate}
    \item The $B_\perp$ resolution is artificially increased by using an interpolation function. This gives no additional information however allows to determine the $B_\perp$ periodicity more precisely in the next step. In the so obtained table for $I\ti{c-SQUID}(B_\perp)$ we retain only the ordinates $I\ti{c}$ and dismiss the equidistant abscissas $(B_\perp)$.
    \item The operation $f_{I\ti{c}}(w)$ acting on the $I\ti{c}$  array with argument $w$  is so  defined:
    \begin{itemize}
        \item a moving average with width $w$ us applied to the list of $I\ti{c}$ values. Here $w$ gives the number of points, that are averaged. A moving average of width $w$ applied to a list $(x_i)$ of length $n$ is defined as: $\mathcal{A}(x_i,w)=\sum_{j=i}^{i+w}x_i/w$ where $i$ runs up to $n+1-w$.
        \item the first numerical derivative of this moving average data is obtained by calculating the difference between each value and the next one in the list. 
        \item $f_{I\ti{c}}(w)$ is then the standard deviation of the numerical derivative list obtained in the previous point. 
    \end{itemize}
    \item the precise period is found as $w_m$ minimizing $f_{I\ti{c}}(w)$. 
    \item The moving average $\mathcal{A}(I\ti{c},w_m)$ gives the background, which is then fitted by a third degree polynomial.
    \item Such polynomial is subtracted from the original $I\ti{c-ref}(B_\perp)$ dataset so the CPR without background remains. 
\end{enumerate}


\section{Modeling of inductive screening effects in the asymmetric SQUID}
\label{app:beta}
An asymmetric SQUID with $I\ti{c-reference}\gg I\ti{c-probe}$ allows to directly probe the CPR of the probe junction when self inductance of the loop is negligible. Below we present an estimate of the inductance parameter $\beta=2IL\ti{l}/\Phi_0$ for the SQUIDs used. 

The model used to derive the CPR for a SQUID with finite self inductance is described following \cite{barone}.
The total supercurrent of a SQUID, $I\ti{SQUID}$, formed by the two junctions a and b with critical currents $I\ti{a}$ and $I\ti{b}$ is given by

\begin{align}
    \label{eq:icsq}I\ti{SQUID}(\phi\ti{e})&=I\ti{a}\text{CPR}\ti{norm}\left(\varphi\ti{a}\right)+I\ti{b}\sin\left(\varphi\ti{b}\right),        
 \end{align}
where junction b has a sinusoidal current phase relation and CPR$\ti{norm}$ is the normalized current phase relation of the probing junction a of arbitrary shape. 
Considering the flux quantization in a SQUID and using Lagrange multipliers, the following relation between the phase drop at junction a, $\varphi\ti{a}$, and the phase drop at junction b, $\varphi\ti{b}$, is derived as:
 \begin{align}
    \label{eq:phib}  \varphi\ti{b}&=\arccos\left(\frac{-1}{I\ti{b}/\left(I\ti{a}\text{CPR}'\ti{norm}\left(\varphi\ti{a}\right)+\beta\ti{b}+\beta\ti{a}\left(I\ti{b}/I\ti{a}\right)\right)}\right).
\end{align} 
Here $\beta_i$ is the self inductance parameter of the respective junction branch.
Pairs $(\varphi\ti{a},\varphi\ti{b})$ satisfying this relation are then used to compute the flux bias applied by the external field:
\begin{align}
  \label{eq:phiext}  \phi\ti{e}&=1/2\pi\cdot\left(\varphi\ti{b}-\varphi\ti{a}+\pi\beta\ti{b}\sin\left(\varphi\ti{b}\right)-\pi\beta\ti{a}\text{CPR}\ti{norm}\left(\varphi\ti{a}\right)\right).
\end{align}





\begin{figure}
    \centering
    \includegraphics[width=0.8\linewidth]{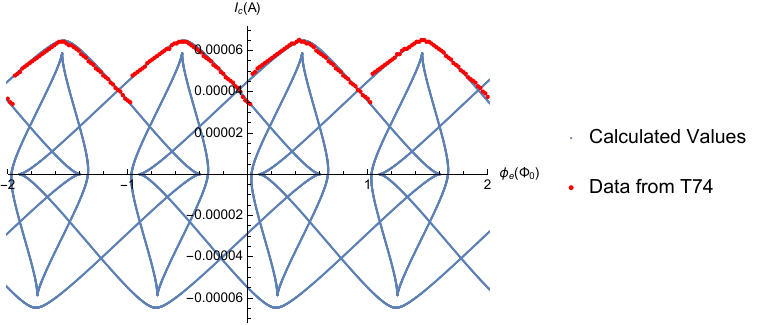}
    \caption{Calculations done for: $\beta\ti{a}=0.636$,  $\beta\ti{b}=0.864$, $I\ti{a}=I\ti{b}=32.38\,$µA. Measurement done at ~$40\,$mK and $\Phi_0$ corresponds to $13.5\,$µT}
    \label{fig:T74-CPR-fit}
\end{figure}

\begin{figure}
    \centering
    \includegraphics[width=0.8\linewidth]{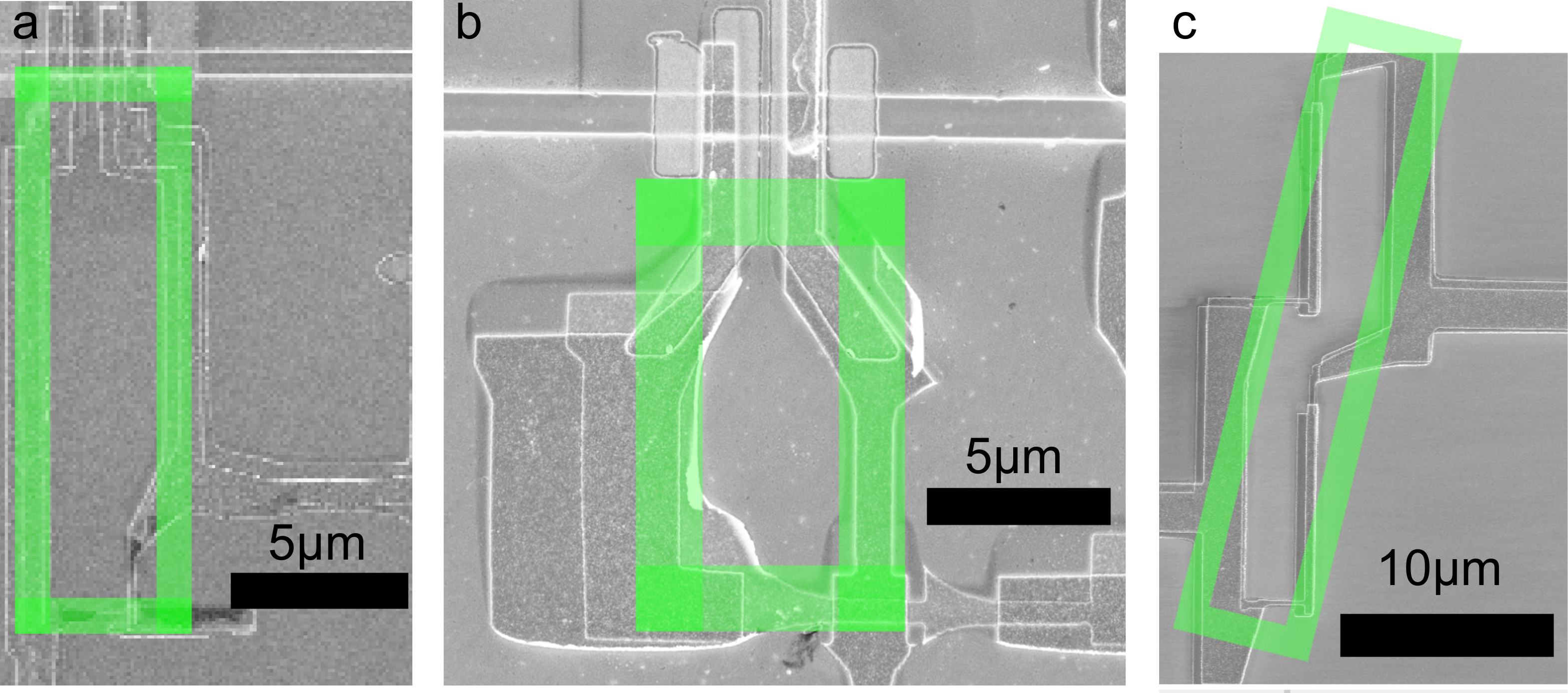}
    \caption{a: A with $l=16\,$µm, $w=5\,$µm, $t=134\,$nm, $h=1\,$µm, b: B with $l=12.3\,$µm, $w=7.3\,$µm, $t=110\,$nm, $h=1.8\,$µm, c: T74 with $l=30\,$µm, $w=6.5\,$µm, $t=134\,$nm, $h=1.4\,$µm. }
    \label{fig:geometric-overview}
\end{figure}

Using \ref{eq:phiext}, we receive the triples $(\varphi\ti{a},\varphi\ti{b},\phi\ti{e})$ from the pairs of allowed $(\varphi\ti{a},\varphi\ti{b})$ calculated before. Using these triples $(\varphi\ti{a}, \varphi\ti{b}, \phi\ti{e})$ in Eq. \ref{eq:icsq} yields $I_\text{SQUID}(\phi\ti{e})$, which can then be compared to experimental data to extract $\beta_i$.
Since the CPR of samples A and B is a priori unknown, and their self-inductance cannot be measured independently, we use a separate symmetric SQUID sample, T74, with identical SIS junctions and a known sinusoidal CPR on each junction. This sample was designed with a sufficiently large $\beta$ to allow accurate quantification of the inductance.
Figure \ref{fig:T74-CPR-fit} shows $I_\text{SQUID}(\varphi\ti{e})$ for T74, with measured data in red and model fits in blue using $\beta\ti{a} = 0.636$, $\beta\ti{b} = 0.864$, and $I_a = I_b = 32.38\,$μA. This corresponds to a total loop inductance of $L_\text{exp,T74} = 47.9\,$pH.
To estimate the loop inductance of A and B, we model each SQUID geometrically as a rectangular loop, as shown in Fig.\ref{fig:geometric-overview} a-c with the rectangle lengths $l\ti{A, B, T74}=16, 12.3, 30\,$µm, the rectangle widths $w\ti{A, B, T74}=5, 7.3, 6.5\,$µm, the thicknesses $t\ti{A, B, T74}=134, 110, 134\,$nm, and the lead widths $h\ti{A, B, T74}=1, 1.8,1.4\,$µm.

The geometric inductance for each SQUID is calculated following \cite{rosa1908self} as
\begin{align}
    L\ti{geo}&=\mu_0/\pi\cdot\left(-0.5(l+w)+2\sqrt{l^2+w^2}+0.447(t+h)+l\ln\left(\frac{2lw}{\left(l+\sqrt{l^2+w^2}\right)(t+h)}\right)+w\ln\left(\frac{2lw}{\left(w+\sqrt{l^2+w^2}\right)(t+h)}\right)\right).
\end{align}
Using the dimensions of the three devices (A, B, T74), this yields $L\ti{geo, A}=22.5\,$pH, $L\ti{geo, B}=18.4\,$pH, and $L\ti{geo, T74}=39.7\,$pH.
Comparing the experimental and geometric values for the inductance of T74 gives $L\ti{exp, T74}/L\ti{geo, T74}=1.21$. We use this factor to predict the inductance of sample A and B from their geometric values, obtaining  $L\ti{pred, A}=27.2\,$pH, $L\ti{pred, B}=22.3\,$pH. \\
A possible explanation for the discrepancy between geometric and experimentally determined inductance include geometrical simplifications in the modeling and the contribution of kinetic inductance. The kinetic inductance can be estimated as
\begin{align}
    L\ti{kin}&=\frac{1}{\sigma_2\omega}=\frac{\hbar}{\sigma\ti{n}\pi\Delta},\\
    L_{\text{kin}\square} &=\frac{\hbar R_\square}{\pi\Delta}
    \label{eq:Lkin}
\end{align}
for $T\ll T\ti{c}$ \cite{tinkham2004introduction}. 
From a single SIS junction - sample T55 (see Fig. \ref{fig:getDelta}) we get $\Delta=325\,$µeV and $\rho\ti{n}=4.0\,$µ$\Omega$cm, see \ref{app:alcharact}. Using \ref{eq:Lkin} we calculate $L\ti{theo}=L\ti{geo}+L\ti{kin}$ we get $L\ti{theo, A}=70.8\,$pH and $L\ti{theo, B}=47.6\,$pH. These values are markedly larger than those of $L\ti{pred}$. 
For further discussion, we denote $L\ti{med}=\sqrt{L\ti{pred}\cdot L\ti{theo}}$ with $L\ti{pred}$ and $L\ti{theo}$ as lower and upper border respectively.\\

Now that we have estimations for the loop inductance, we want to check how large they impact the determined $\tau$. This is necessary since the self inductance increases the skewness of the probed CPR in a similar matter as $\tau$ does.

Using \ref{eq:phiext} and \ref{eq:phib} we model $I\ti{SQUID}(\phi\ti{e})$ for A with $I\ti{a}=250\,$nA, $I\ti{b}=7\,$µA and B with $I\ti{a}=250\,$nA, $I\ti{b}=24.5\,$µA. This is done for $L\ti{l}= L\ti{pred}, L\ti{med}, L\ti{theo}$ with
\begin{align}
    \label{eq:ballisticNorm}\text{CPR}\ti{norm}(\varphi)&=\frac{1}{\text{Amp}}\cdot\frac{\sin(\varphi)}{2\sqrt{1-\tau\sin^2(\varphi/2)}},\\
    \label{eq:ampnorm}\text{Amp}&=\frac{\sin\left(2\arccos\left(\sqrt{1-1/\tau+\sqrt{1-\tau}/\tau}\right)\right)}{2(1-\tau)^{1/4}}.
\end{align}
Here the right hand side of \ref{eq:ballisticNorm} is derived from the CPR of a ballistic junction in the low temperature regime. The initial $\tau$ however is picked such that fitting the resulting $I\ti{SQUID}(\phi\ti{e})$ gives $\tau\ti{fit}=0.947, 0.84$ for A, B respectively, see Fig.~2 of the main text. The resulting corrected values are the following $\tau\ti{med, A}=0.941+0.004-0.007$ and $\tau\ti{med, B}=0.821+0.006-0.010$.

\subsection{Impact of the self inductance on
$\varphi_0$}
To discuss the impact of the self inductance onto $\varphi_0$, we consider the flux quantization condition
\begin{align}
    2\pi n=\varphi\ti{a}-\varphi\ti{b} +2\pi\phi\ti{e}+\beta\ti{a}\sin\varphi\ti{a}-\beta\ti{b}\sin\varphi\ti{b}.
\end{align}
For a strongly asymmetric SQUID with $I\ti{b}\gg I\ti{a}$, $\varphi\ti{b}=\pi/2$ in order to maximize the critical current of the total SQUID, allowing to simplify our quantization condition to 
\begin{align}
  \varphi\ti{a} =\varphi\ti{b} -2\pi\phi\ti{e} -\beta\ti{a}\sin\varphi\ti{a} +\beta\ti{b}.
\end{align}
We see that in this asymmetric case, the self inductance of the reference junction path gives a constant offset to the phase of the probe junction, given by $\beta\ti{b}$. Inverting the SQUID bias current will consequently lead to  $\varphi\ti{b}=-\pi/2$ and therefore also invert the constant phase offset to $-\beta\ti{b}$. 
This observation both justifies and shows the necessity of determining $\varphi_0$, by evaluating the CPR for positive and negative SQUID bias currents, and then taking the mean value of $\varphi_0$ for each bias direction. In agreement with the expected behavior of $\varphi_0(B\ti{inplane})$, this correction yields $\varphi_0(B\ti{inplane}=0)\approx0$. 
For Fig.~2b of the main text this correction step was used to overlay the CPRs of both bias directions (together with application of a two point moving average), while the other CPRs stem from a single bias direction.

\section{Determination of the in plane angle between wire orientation and magnetic field}
\begin{figure}
    \centering
    \includegraphics[width=0.6\linewidth]{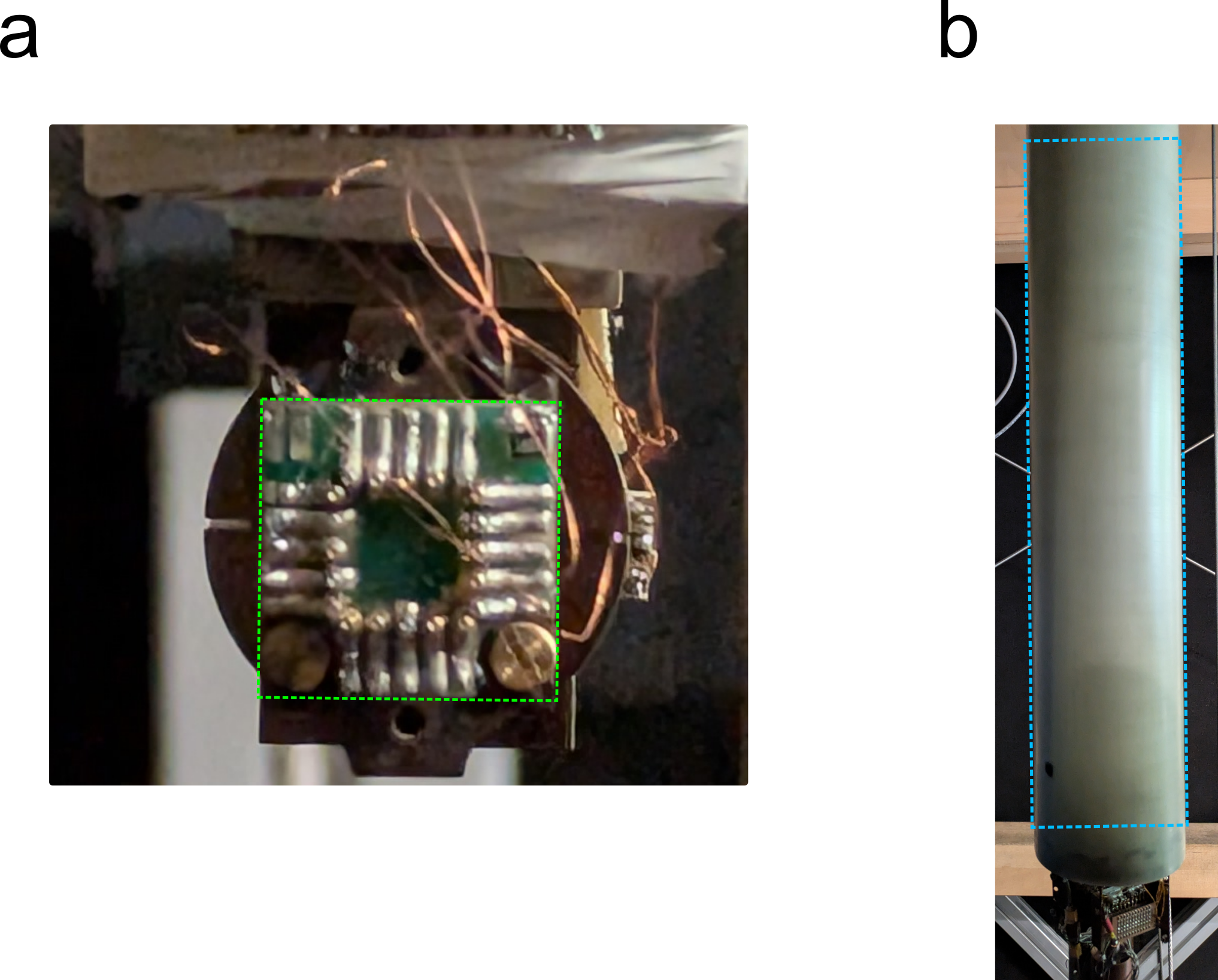}
    \caption{Images of (a) the sample carrier (mounted on a piezo rotator, whose round Cu base is visible) and (b) cryostat insert. Pictures are taken with a camera with internal level determination based on an inertial measurement unit. This ensures that pictures are aligned with the gravitational field. The $x$-$z$  reference frame for the sample carrier and for the cryostat are here represented by the green and blue rectangle, respectively.}
    \label{fig:alignment-bfield}
\end{figure}

Since the in-plane angle between the wire and the magnetic field plays a major role, see Fig.~3b of the main text, in this section we discuss how we determined this angle and, importantly, the \textit{error} associated to the angle determination.
There are several sources of in-plane misalignment that need to be considered:
\begin{itemize}
    \item The misalignment of the chip with respect to the chip-carrier, and thus to the cryostat axis.
    \item The error induced by the rotator of the sample (slip-stick piezo rotator), in particular the drift associated by multiple sweeps of the angle at low temperature.
    \item The error induced by misalignment of magnet and insert.
\end{itemize}
\textbf{Chip with respect to carrier:} This is the most straightforward angle to be determined,  since the orientation of the wire can be determined from a low-magnification optical microscope image that also shows the orientation of the chip carrier. The image readout is done digitally. This gives an angle correction of $1\text{°}\pm 0.3°$°. The correction is applied to the rotational values, while the uncertainty of $\pm0.3$°  contributes to the total error.
\textbf{Rotator error:} To estimate this error, the rotator was initially (before cooldown) aligned such that one side of the chip carrier was parallel to the direction of the gravitational force. During cooldown the rotator was then  rotated back and forth multiple times. After the cooldown, the alignment with the gravity was checked again. The difference of $0.7$° corresponds to the typical error due to drifts at low temperature. This error is then added to the total error see Fig. \ref{fig:alignment-bfield} a.
\textbf{Magnet with respect to the cryostat axis:} This error source is separated into alignment of the cryostat axis and alignment of the magnet coil, both considered with respect to the gravity field. The former was determined by a using a camera equipped with an internal inertial measurement unit, which allows one to determine the orientation of the picture with respect to gravity, see Fig.~\ref{fig:alignment-bfield} b. This gives us an error of $0.6$°.  The remaining misalignment of the magnet with respect to gravity field is more difficult to determine directly. However, we can estimate that an error larger than 1.5 degrees would prevent proper insertion of the cryostat into the magnet core.  
Therefore the total error taken as the simple sum of the individual errors (maximum error, pessimistic scenario) is about 3°, as specified in the main text. For comparison, a more realistic quadrature sum of the errors would provide an error of 1.8°, dominated by the magnet-to-gravity alignment uncertainty.

As an independent check, we also tried to determine the angular offset from the in-plane field dependence of the Fraunhofer pattern of the tunnel junction. 
The resulting reference angle deviates by 3° from the value obtained with the geometrical method discussed above.
However, the \textit{error bar} was much larger, namely $\pm$9°. Therefore, the zero angle used in the analysis was determined using the geometrical method discussed above. The resulting $\pm$3° error bar associated with that method is sufficiently large to include the value obtained from the SIS Fraunhofer pattern calibration. We are therefore confident that the zero angle in the graph in Fig.~3b is determined with an accuracy of $\pm$3°, which is much smaller than the reported 19° of the spin-to-momentum angle. 

\section{Determination of $\Delta^*$ and channel number from $T$ dependent measurements}
Here, we summarize the determination of the junction parameters from the measurement shown in Fig.~2 of the main text:
\begin{itemize}
    \item At 30-40mK, where $\Delta^*\ll k\ti{B}T$, $I(\varphi)=I_0\tau\sin(\varphi)/\left(2\sqrt{1-\tau\sin^2(\varphi/2)}\right)$ \cite{RMPGolubov}. This allows to extract $\tau$ in this temperature regime directly from the skewness and to use it as a fixed parameter for subsequent fits.
    \item Using the finite temperature expression $I(\varphi)=I_0\tau\sin(\varphi)\tanh\left[\frac{\Delta^*}{k\ti{B}T}\sqrt{1-\tau\sin^2(\varphi/2)}\right]/\left(2\sqrt{1-\tau\sin^2(\varphi/2)}\right)$, with $I_0=n\ti{channels}e\Delta^*/\hbar$ the CPRs in the temperature range 100-700~mK are fitted. 
    \item Since $n\ti{channels}$ and $\Delta^*$ both enter the amplitude linearly, it is not possible to decouple them by fitting a single CPR at a given $T$. Therefore $n\ti{channels}\epsilon\mathds{N}$ was treated as a fixed parameter during the fit procedure, leaving $\Delta^*$ as the only free parameter. 
    \item Following this approach, the CPR at each temperature was fitted with a predefined value of $n\ti{channels}$. \item To receive the correct value for $n\ti{channels}$, the overall variance for all fits with one predefined $n\ti{channels}$ was analyzed.
\end{itemize}

\end{document}